\documentclass[11pt,a4paper, linktocpage,colorlinks=true, linkcolor=bluscuro,urlcolor=bluscuro,citecolor=bluscuro]{article}
\usepackage{jheppub}
\usepackage{graphicx}
\usepackage{bm}
\usepackage{xcolor}
\usepackage{booktabs}
\PassOptionsToPackage{breaklinks}{hyperref}
\definecolor{bluscuro}{rgb}{0.15, 0.2, .85}
\usepackage{amsmath, amssymb}
\usepackage{slashed}
\usepackage{latexsym}
\usepackage{braket}
\usepackage[export]{adjustbox}
\usepackage{color}
\usepackage{multirow}
\usepackage{xspace}
\usepackage{comment}
\usepackage[caption=false]{subfig}
\usepackage{cancel}
\usepackage{url}
\usepackage{nicematrix}

\title{Probing the multipolar structure of Myers-Perry black holes with scattering amplitudes}

\author[a, b]{Massimo Bianchi,}
\author[c, d]{Claudio Gambino,}
\author[d]{Fabio Riccioni,}
\author[a]{Vincenzo Zevola}
\affiliation[a]{Dipartimento di Fisica,  Universit\`a di Roma ``Tor Vergata"}
\affiliation[b]{Sezione INFN Roma2, Via della Ricerca Scientifica 1, 00133, Roma, Italy}
\affiliation[c]{Dipartimento di Fisica,  Universit\`a di Roma ``La Sapienza"}
\affiliation[d]{Sezione INFN Roma1, Piazzale Aldo Moro, 00184, Roma, Italy}
\emailAdd{mbianchi@roma2.infn.it, claudio.gambino@uniroma1.it, fabio.riccioni@roma1.infn.it}
\abstract{We discuss the scattering of massive scalar probes off Myers-Perry black holes in the Kerr-Schild  gauge. Extending the analysis performed recently for Kerr(-Newman) black holes, we show that the Kerr-Schild gauge allows to write down the tree-level scattering amplitude for Myers-Perry black holes in analytic form. For generic values of the angular momenta, Myers-Perry solutions have a richer multipolar structure compared to their four-dimensional counterparts, because they are  characterized by the presence of stress multipoles, together with the more familiar mass and current multipoles. By focusing on the five-dimensional case, we derive the leading eikonal phase from the scattering amplitude and we give an explicit expression for two limiting scenari, namely when the two angular momenta are the same, so that the mass multipoles vanish but still the solution has a non-vanishing stress quadrupole and a current dipole, and when one of the two angular momenta is zero, and correspondingly the stress multipoles vanish similar to the Kerr case.}
\keywords{Scattering Amplitudes, Black Holes, Multipoles}

\begin{document}
\maketitle

\section{Introduction}
Since the pioneering investigation by Rutherford of atomic and nuclear structures, scattering processes have been systematically used to probe the internal structure of composite or `complex' objects: from atoms to nuclei, from protons to... black holes (BHs). The latter look like unique `elementary' objects in classical General Relativity (GR), due to the no-hair theorem, but present many paradoxical features as soon as one includes quantum effects~\cite{Hawking:1973uf}. In particular no commonly accepted explanation for the origin of the BH entropy in terms of explicit micro-states is available, despite many old and recent attempts~\cite{Bardeen:1973gs, Jacobson:1993vj, Strominger:1996sh, Bena:2007kg, Haco:2018ske}.

While in four spacetime dimensions it is well known that the Kerr metric is the unique black hole solution in GR~\cite{Morisawa:2004tc}, in higher dimensions BH uniqueness is lost\footnote{Attempts to prove some form of a uniqueness theorem for BHs in higher dimensions were made by requiring the stability of the solutions~\cite{Kol:2002dr}.} (see~\cite{Emparan:2008eg} for a recent review), in that several horizon topologies are possible, such as black rings with a $S^2\times S^1$ horizon~\cite{Emparan:2001wn} or black saturns~\cite{Elvang:2007rd} in $D=5$ (see~\cite{Hollands:2006rj} for solutions with at least one Killing vector that generates rotations), several higher-form fields might be present and several components of the angular momentum can be turned on. A particularly interesting class of rotating BHs with arbitrary angular momentum in higher dimensions have been constructed by Myers and Perry (MP)~\cite{Myers:1986un}. MP BHs `naturally' generalize Kerr BHs~\cite{Kerr:1963ud} in that the topology of the horizon is a sphere~\cite{Morisawa:2004tc}. Moreover similarly to the Kerr BHs, they admit a Kerr-Schild (KS) gauge~\cite{Monteiro:2014cda} that drastically simplifies the study of scattering processes. 

Such gauge, at least for scalar probes, makes possible to write scattering amplitudes in the probe limit only using a tri-linear vertex. Thanks to miraculous cancellations between the measure and the harmonic function representing the gravitational potential, computing tree-level elastic amplitudes becomes as `straightforward' as for Kerr(-Newman) BHs~\cite{Bianchi:2023lrg}.
With some ingenuity, in the  KS gauge, one can compute the Fourier transform (FT) of the metric in hyper-spherical coordinates. The result can be expressed in terms of Bessel functions of the variable $u_a$ that combines the transferred momentum $q$  with  the `spin' matrix $S$, as we will see. Imposing mass-shell conditions one can derive a Rutherford-like formula for low-mass probes $m\ll M$. As for Kerr BHs spin produces quantitatively small corrections to the non-rotating Schwarzschild(-Tangherlini) case. Not surprisingly, the very same Bessel functions that appear in scattering amplitudes appear in the  multipolar structure of MP BHs, recently studied in momentum space~\cite{Bianchi:2024shc}. In this respect it is worth to recall that a new class of multipoles appear in higher dimensions that were dubbed `stress multipoles' by  P.~Pani and two of the present authors (C.~G. and F.~R.)~\cite{Gambino:2024uge} (see also~\cite{Mayerson:2022ekj, Heynen:2023sin, Amalberti:2023ohj}), generalizing the four dimensional analysis pioneered in~\cite{Geroch:1970cc, Geroch:1970cd, Hansen:1974zz,Thorne:1980ru, Gursel1983}. 

In fact one can show that the most general stress-energy tensor of a compact object with no other intrinsic properties other than mass and angular momentum can be parameterized in terms of structure functions encoding its multipolar structure~\cite{Bianchi:2024shc}. These are the gravitational analogue of the form factors and structure functions introduced to analyze (deep inelastic) scattering off composite particles like protons and neutrons.

For scalar probes one can in principle compute higher-loop amplitudes and isolate the terms that survive in the classical limit. These can be packaged in a gauge-invariant combination, usually the eikonal phase, related to the phase-shifts in the partial wave basis that diagonalizes the S-matrix. For large impact parameters, \textit{i.e.} small deflections, the eikonal phase can be computed and shown to exponentiate (see {\it e.g.}~\cite{DiVecchia:2023frv} for a recent review). Radiation losses and other self-force effects can be included in the Extreme-Mass-Ratio-Inspiral regime $m\ll M$. Relying on modern methods\footnote{See {\it e.g.}~\cite{Travaglini:2022uwo, Buonanno:2022pgc, Adamo:2022dcm} for recent reviews.},  one can reconstruct the full (gravitational) wave-form and use it as a test of GR in higher dimensions by carefully comparing it with the one predicted by probe scattering off other Exotic Compact Object (ECO), such as fuzzballs. Eventually, one would like to answer the pressing question: {\it ``Can we tell a BH from a fuzzball or another ECO?''} Probably we are not yet there, but future observations of gravitational wave signals from binary mergers or close encounters {\it of the third kind}~\cite{LIGOScientific:2016aoc, LIGOScientific:2019fpa, EventHorizonTelescope:2019dse, EventHorizonTelescope:2022xqj} may help discriminating BHs from other very compact objects~\cite{Raposo:2018xkf,Raposo:2020yjy,Bena:2020see,Bianchi:2020bxa,Bena:2020uup,Bianchi:2020miz,Fransen:2022jtw}.

One should not forget that BH mimickers~\cite{Cardoso:2019rvt, Gambino:2025xdr} can exist already in $D=4$. An example {\it in vacuum} with multipole moments identical to those of Kerr was given in~\cite{Bonga:2021ouq} in the post-Newtonian limit and then generalized in~\cite{Gambino:2025xdr} in the relativistic case, while allowing for the presence of electro-magnetic field the Kerr-Newman BH~\cite{Newman:1965tw} has the same gravitational multipole moments as the Kerr BH~\cite{Sotiriou:2004ud}, even though the bound on the angular momentum gets modified from $a=J/M\le M$ to $a\le \sqrt{M^2-Q^2}$ for charged BHs. 

In higher dimensions different horizon topologies are allowed and BH solutions in $D\ge 5$ are known  with no bound on their spin (components). Similarly over-rotating smooth horizonless objects are known to exist~\cite{Jejjala:2005yu} that are however affected by ergoregion~\cite{Cardoso:2005gj} and charge instabilities~\cite{Bianchi:2023rlt, Cipriani:2024ygw}. Gregory-Laflamme instability can also occur~\cite{Gregory:1993vy} but, to the best of our knowledge, thanks to the spherical topology of the horizon, they do not affect MP BHs, that are the subject of our present analysis.

The plan of the paper is as follows. In Section~\ref{sec:KerrRev} we will briefly review Kerr(-Newman) BHs in the KS gauge and describe how to compute scattering amplitudes and cross-sections for massive (neutral) scalar probes. We then move to consider MP BHs in $D=d+1$ dimensions in Section~\ref{sec:MP}. We will work in the KS gauge and highlight the differences between $D$ even and $D$ odd.
In Section~\ref{sec:CrossSection} we study the scattering of a massive scalar on MP BHs in arbitrary $D$ dimensions. 
Comparison with the multipolar structure of BHs will be studied in Section~\ref{sec:MultiRev}.
In Section~\ref{sec:D5Case} we will focus on the $D=4+1$ case and explicitly compute the leading eikonal phase for special choices of the angular momenta, {\it viz.} $a_1=\pm a_2=a$ and $a_2=0$, and of the kinematics. 
Finally Section~\ref{sec:Conclusion} contains our conclusions and outlook.

Relevant but cumbersome formulae will be given in the Appendices. In Appendix~\ref{app:ExpCalc} we compute the FT of the metric in arbitrary  $D=d+1$, while in Appendix~\ref{app:Jacobian} we compute the Jacobian for the transformation from cartesian to poly-spherical coordinates in $d$ dimensions, making a distinct analysis for even and odd $d=D-1$.

\textbf{Conventions} 
As in~\cite{Bianchi:2023lrg} we use mostly minus conventions \textit{i.e.} $p^2=m^2$, as usual we set $c=\hbar=1$. Moreover, we consider spacetime with an arbitrary number $D=d+1$ dimensions and use $D$ and $d$ interchangeably for convenience. Moreover the combination $\alpha= \frac{D-3}{2}$ will appear ubiquitously in the paper.

\section{Kerr(-Newman) BHs and {KS} gauge: a lightning review}\label{sec:KerrRev}
Let us briefly review the analysis performed in~\cite{Bianchi:2023lrg} for the scattering of a scalar massive probe off a Kerr(-Newman) black hole~\cite{Kerr:1963ud}. In particular, we recall the {KS} gauge for Kerr(-Newman) BHs and its properties and sketch how to perform the FT of the metric that appears in the computation of scattering amplitudes.

The Kerr(-Newman) metric in {KS} gauge reads
\begin{equation}
 \label{eq:forma della metrica di {KS}}
    g_{\mu\nu}=\eta_{\mu\nu}+h_{\mu\nu}=\eta_{\mu\nu}+\Phi_G K_\mu K_\nu\ ,
\end{equation}
where the gravitational potential is given by
\begin{equation}
\Phi_G=- \frac{G(2Mr-Q^2)}{r^2+a^2\cos^2{\theta}}\ ,
\end{equation}
with $M$ the mass, $Q$ the charge and $J=Ma$ the angular momentum of the BH, and 
\begin{equation}
    K_\mu=\Bigg(1,\frac{rx+ay}{r^2+a^2},\frac{ry-ax}{r^2+a^2},\frac{z}{r}\Bigg)
\end{equation}
is a null vector since the radial coordinate $r$ is defined via
\begin{equation}
    \frac{x^2+y^2}{r^2+a^2}+\frac{z^2}{r^2}=1\ .
\end{equation} 
One of the main properties of the {KS} gauge is that the inverse metric is still linear in the gravitational constant
\begin{equation}
    g^{\mu\nu} = \eta^{\mu\nu} - h^{\mu\nu}=\eta^{\mu\nu}-\Phi_G K^\mu K^\nu\ .
\end{equation}
Henceforth we will restrict to the neutral case and set $Q=0$, even though our results can be generalized to charged BHs. Note however that the five-dimensional Cheung-Cvetic-Lu-Pope (CCLP) charged-rotating BHs~\cite{Chong:2005hr} do not admit a KS gauge. 

As is well known the singularity (or `ringularity') of a Kerr BH is a ring of radius $a$ in the equatorial plane, corresponding to $r=0$, \textit{i.e.} $z=0$ (${\theta =\pi/2}$) and ${x^2+y^2=a^2}$. Moreover,
Kerr BHs are covered by two horizons with spherical topology with radii  ${r_H^{\pm} = M \pm \sqrt{M^2-a^2}}$. The condition $a\le M$ has to be imposed to avoid naked singularities. 

Our interest here is the scattering of a massive scalar probe off a Kerr BH. In the KS gauge, the tree-level scattering amplitude is fully encoded in a tri-linear vertex 
\begin{equation}
\label{eq:definizione ampiezza di diffusione}
    2\pi\delta(q_0)i\mathcal{M}(p,p^\prime,\vec{q}\,)=\frac{i}{2}\widetilde{h}^{\mu\nu}(q)\widetilde{T}^{\rm matt}_{\mu\nu}(p,p^\prime)=\frac{i}{2}2\pi\delta(q_0)\widetilde{h}^{\mu\nu}(\vec{q}\,)\widetilde{T}^{\rm matt}_{\mu\nu}(p,p^\prime)\ ,
\end{equation}
where $\tilde{h}_{\mu\nu}$ and $\widetilde{T}_{\mu\nu}^{\rm matt}$ are respectively the FT of the metric perturbation and probe energy-momentum tensor and $q$ is the transferred momentum. 

 Therefore, the time independent FT of the stationary Kerr metric 
\begin{equation}
\widetilde{h}_{\mu\nu}(q)=2\pi\delta(q_0)\int d^3\vec{x} e^{-i\vec{q}\cdot \vec{x}}\Phi_G(\vec{x})K_\mu(\vec{x})K_\nu(\vec{x})\ ,
\end{equation}
is needed. Quite remarkably the overall factor in the  volume element in oblate spherical coordinates
\begin{equation}
    d^3\vec{x}=dxdydz=(r^2+a^2\cos^2{\theta})\sin{\theta}drd\theta d\phi\ ,
\end{equation}
exactly and crucially cancels the denominator in $\Phi_G$. In performing the FT, we find it convenient to introduce
\begin{equation}
 \vec{u}=(q_x\sqrt{r^2+a^2},q_y\sqrt{r^2+a^2},q_zr)  \quad \mathrm{and} \quad  \vec{n}=(\sin{\theta}\cos{\phi},\sin{\theta}\sin{\phi},\cos{\theta})\ , 
\end{equation}
so that $\vec{n}\cdot\vec{n}=1$ and 
\begin{equation}
u=|\vec{u}|=\sqrt{r^2q^2+a^2q_\perp^2} \ , \qquad  q^2=|\vec{q}\, |^2=q_x^2+q_y^2+q_z^2=q_\perp^2+q_z^2\ ,
\end{equation}
and write
\begin{equation}
\vec{q}\cdot\vec{x}=(q_x\cos{\phi}+q_y\sin{\phi})\sin{\theta}\sqrt{r^2+a^2}+q_zr\cos{\theta} = \vec{u}\cdot\vec{n} \ .
\end{equation}
The angular integral then takes the form of a Bessel integral 
\begin{equation}
    \frac{1}{4\pi} \int d\Omega e^{-i\vec{u}\cdot\vec{n}}=\frac{\sin{u}}{u}=j_0(u)\ ,
\end{equation}
where $j_0(u)$ is the spherical Bessel function. Replacing $ \vec{x}=i\partial_{\vec{q}}$  inside the FT, so that
\begin{equation}\label{eq:K nella rappr degli impulsi}
K_\mu(r,\vec{x}=i\partial_{\vec{q}})=\Bigg(1,i\frac{r\partial_{q_x}+a\partial_{q_y}}{r^2+a^2},i\frac{r\partial_{q_y}-a\partial_{q_x}}{r^2+a^2},i\frac{\partial_{q_z}}{r}\Bigg)\ ,
\end{equation}
one gets 
\begin{equation}\label{eq:hTransformD4}
\widetilde{h}_{\mu\nu}(\vec{q}\,)=-8\pi G M\int_0^\infty r dr K_\mu(r,\vec{x}=i\partial_{\vec{q}})K_\nu(r,\vec{x}=i\partial_{\vec{q}})j_0(u)\ .
\end{equation}
Changing integration variable from $r$ to $u$, where
\begin{equation}
|\vec{q}\, |^2r^2=u^2-a^2q_\perp^2\ , \qquad |\vec{q}\, |^2(r^2+a^2)=u^2+a^2q_z^2\ , \qquad   rdr=\frac{udu}{|\vec{q}\, |^2} \ ,
\end{equation}
and taking into account that $K_\mu$ acting on functions of $u$ produces
\begin{equation}
\label{eq:azione di K su una fun di u}
    K_\mu F(u)=\Bigg(1,i\Big(r(u)q_x+aq_y\Big),
    i\Big(r(u)q_y-aq_x\Big),ir(u)q_z\Bigg)\frac{1}{u}\frac{d}{du}F(u)\ ,
\end{equation}
it is possible to evaluate Eq. \eqref{eq:hTransformD4} by means of the master integrals
\begin{equation}
    {\cal J}_n=\int_{q_\perp a}^\infty du\frac{u^{1-n}j_n(u)}{\sqrt{u^2-q_\perp^2a^2}}=\frac{\pi}{2}\frac{J_n(q_\perp a)}{(q_\perp a)^n}\ ,
\end{equation}
where $J_n$ are Bessel functions of the first kind.

The explicit expression for the FTs of the metric components in the (neutral) Kerr case read
\begin{align}
    \widetilde{h}_{00}(\vec{q}\,)&=- \frac{8\pi GM}{|\vec{q}\, |^2}\cos{|\vec{a}\times\vec{q}\,|}\ ,\\
     \widetilde{h}_{0i}(\vec{q}\,)&=- i 8\pi GM \left[\frac{ (\vec{a}\times\vec{q}\,)_i}{|\vec{q}\, |^2}j_0(|\vec{q}\times\vec{a}|)-\frac{q_i}{|\vec{q}\, |^3}\frac{\pi}{2}J_0(|\vec{q}\times\vec{a}|)\right]\ ,\\
    \widetilde{h}_{ij}(\vec{q}\,)&=- 8\pi GM \left[\frac{j_0(|\vec{a}\times\vec{q}\,|)}{|\vec{q}\, |^2}\Big(\delta_{ij}-2\frac{q_iq_j}{|\vec{q}\, |^2}\Big)+\frac{1}{|\vec{q}\, |^3}\frac{\pi}{2}\frac{J_1(|\vec{a}\times\vec{q}\,|)}{|\vec{a}\times\vec{q}\,|}(q_i(\vec{a}\times\vec{q}\,)_j+q_j(\vec{a}\times\vec{q}\,)_i) \right.\nonumber\\ 
    &\left. -\frac{1}{|\vec{q}\, |^2}\frac{j_1(|\vec{a}\times\vec{q}\,|)}{|\vec{a}\times\vec{q}\,|}(\vec{a}\times\vec{q}\,)_i(\vec{a}\times\vec{q}\,)_j \right]\ .
\end{align}
Inserting in \eqref{eq:definizione ampiezza di diffusione} the energy-momentum tensor of a massive scalar probe in momentum space
\begin{equation}
    \widetilde{T}^\phi_{\mu\nu}(p,p^\prime)=p_\mu p_\nu^\prime+p_\nu p_\mu^\prime-\eta_{\mu\nu}(p\cdot p^\prime-m^2)\ ,
\end{equation}
with $p^\mu=(E,\vec{p}\,)$ and ${p}'^\mu=(E,\vec{p}\, ')$ and taking into account that $h_{\mu\nu}$ is traceless, one finds
\begin{equation}
    i\mathcal{M}=i\widetilde{h}^{\mu\nu}(q)p_\mu p'_\nu=-i\Bigg(\widetilde{h}_{00}EE^\prime+\widetilde{h}_{0i}(E {p'}^i+E' {p}^i)+\widetilde{h}_{ij}{p^i}{p'}^j\Bigg) \ .
\end{equation}
Assembling the terms and using $E=E^\prime$ (since $q_0=0$) one gets the `off-shell' amplitude
\begin{equation}
\begin{aligned}
    i\mathcal{M}(p,p^\prime,\vec{q}\,)&= i\frac{8\pi GM}{|\vec{q}\, |^2}\Bigg[\cos({|\vec{a}\times\vec{q}\,|})E^2 +iE\Big((\vec{a}\times\vec{q}\,)\cdot (\vec{p}+\vec{p^\prime})j_0(|\vec{q}\times\vec{a}|)-\frac{\vec{q}\cdot(\vec{p}+\vec{p^\prime})}{|\vec{q}|}\frac{\pi}{2}J_0(|\vec{q}\times\vec{a}|)\Big) \\
    &+j_0(|\vec{a}\times\vec{q}\,|)\Big(\vec{p}\cdot \vec{p^\prime}-2\frac{\vec{q}\cdot\vec{p}\, \vec{q}\cdot\vec{p^\prime}}{|\vec{q}|^2}\Big) -\frac{j_1(|\vec{a}\times\vec{q}\,|)}{|\vec{a}\times\vec{q}\,|}(\vec{a}\times\vec{q}\,)\cdot \vec{p}\, (\vec{a}\times\vec{q}\,)\cdot \vec{p^\prime} \\
    &+\frac{1}{|\vec{q}\,|}\frac{\pi}{2}\frac{J_1(|\vec{a}\times\vec{q}\,|)}{|\vec{a}\times\vec{q}\,|}\Big(\vec{q}\cdot\vec{p}\,(\vec{a}\times\vec{q}\,)\cdot\vec{p^\prime}+\vec{q}\cdot\vec{p^\prime}(\vec{a}\times\vec{q}\,)\cdot \vec{p}\Big)\Bigg]\ .
\end{aligned}
\end{equation}
Setting $P= p + p'$ and recalling that on-shell ${q\cdot(p+p') = q{\cdot}{P}=-\vec{q}{\cdot}{\vec{P}}=0}$ one gets 
\begin{align}
        i\mathcal{M}_{on-shell}=& i\frac{8\pi GM}{|\vec{q}|^2}\Bigg\{\cos(|\vec{a}\times\vec{q}\,|)\Bigg[E^2+\frac{1}{4}\frac{\vec{a}\times\vec{q}\cdot\vec{{P}}}{|\vec{a}\times\vec{q}\,|^2}\:\Bigg]\nonumber \\
        &+\frac{\sin{|\vec{a}\times\vec{q}\,|}}{|\vec{a}\times\vec{q}\,|}\Big[|\vec{p}\,|^2-\frac{1}{4}\frac{(\vec{a}\times\vec{q}\cdot \vec{{P}}\:)^2}{|\vec{a}\times\vec{q}\,|^2}+iE\,\vec{a}\times\vec{q}\cdot\vec{{P}}\,\Big]\Bigg\}\ .
\end{align}
Fermi golden rule allows to derive the differential cross section 
\begin{equation}
    d\sigma=\frac{1}{2Ev}2\pi\delta(E-E^\prime)|\mathcal{M}_{on-shell}|^2\frac{d^3\vec{p}\, ^\prime}{(2\pi)^32E^\prime}\ .
\end{equation}
Integrating over $E'$ one finds
\begin{equation}
\label{eq:sezione d'urto formula}
    \frac{d\sigma}{d\Omega}=\frac{1}{16\pi^2}|\mathcal{M}_{on-shell}|^2\ .
\end{equation}
Factoring out $E^2$ and setting $\frac{{\vec{P}}}{E}\approx\frac{2\vec{p}}{E}=2\vec{v}$ yields a Rutherford-like formula  
 \begin{equation}
    \frac{d\sigma}{d\Omega}\Big|_{\text{Kerr}}=\frac{G^2M^2}{4v^4\sin^4{(\vartheta/2)}}\Bigg|\cos{|\vec{a}\times\vec{q}\,|}\Big(1+\frac{(\vec{a}\times\vec{q}\cdot\vec{v}\,)^2}{|\vec{a}\times\vec{q}\,|^2}\Big)
    +\frac{\sin{|\vec{a}\times\vec{q}\,|}}{|\vec{a}\times\vec{q}\,|}\Big(v^2-\frac{(\vec{a}\times\vec{q}\cdot\vec{v}\,)^2}{|\vec{a}\times\vec{q}\,|^2}+2i\vec{a}\times\vec{q}\cdot\vec{v}\Big)\Bigg|^2\ .
\end{equation}
where $\vartheta$ is the scattering angle related to $|\vec{q}\, |^2 = 4|\vec{p}\, |^2 \sin^2(\vartheta/2)$. 
For non-rotating ($i.e.$ Schwarzschild) BHs one finds the well-known gravitational Rutherford formula
\begin{equation}
    \frac{d\sigma}{d\Omega}\Big|_{\rm Schw.}=\frac{G^2M^2(1+v^2)^2}{4v^{4}\sin^4{(\vartheta/2)}}\ .
\end{equation}

\section{Myers-Perry BHs in D=d+1 dimensions}\label{sec:MP}

We will now turn our attention on rotating solutions in higher dimensions. In particular, we will focus on MP BHs~\cite{Myers:1986un}, that are neutral and admit an event horizon with spherical topology. We will write the metric of MP BHs in the KS gauge~\cite{Monteiro:2014cda} and compute its FT using similar techniques and tricks as for Kerr(-Newman) BHs in $D=4$~\cite{Bianchi:2023lrg}. One has to distinguish from the very beginning the case of $D$ even ($d$ odd) from the case $D$ odd ($d$ even). 

\subsection{MP BH in $D=2n+1$}
In poly-spherical coordinates $(r, \phi_k, \mu_k)$,  with $k=1,..., n$ and $r\in [0,+\infty)$, $\phi_k\in[0,2\pi]$ and $\mu_k\in [0,1]$, subject to the condition
\begin{equation}
\label{eq:cond sui mui}
    \sum_{k=1}^n\mu_k^2=1 \ , 
\end{equation}
the MP metric in $D=2n+1$ can be written as
\begin{equation}
    ds^2=dt^2-\frac{\widehat{M} r^2}{\Pi F}\Bigg(dt+\sum_{k=1}^n a_k\mu_k^2d\phi_k\Bigg)^2-\frac{\Pi F}{\Pi-\widehat{M} r^2}dr^2 \\
    -\sum_{k=1}^n(r^2+a_k^2)(d\mu_k^2+\mu_k^2d\phi_k^2) \ ,
\end{equation}
where $\widehat{M}$ is a `mass parameter', related to the physical BH mass by
\begin{equation}
   \widehat{M} = \frac{16\pi G M } {(D-2) \Omega_{D-2}} \ ,
\end{equation}
with 
\begin{equation}
    \Omega_{D-2}=\frac{2\pi^{\frac{D-1}{2}}}{\Gamma(\frac{D-1}{2})}
\end{equation}
the surface of the unit $(D-2)$-sphere and $a_k$, the MP angular momentum parameters, that  are related to the angular momentum tensor $J$ through\footnote{The parameters $a_k$ are related to the parameters $\mathfrak{a}_k$ used in~\cite{Bianchi:2024shc} by $\mathfrak{a}_k=J_k/M$.}
\begin{equation}\label{eq:DefSpinParam}
    a_k=\frac{D-2}{2} \frac{J_k}{M}\ ,
\end{equation}
where the angular momentum (`spin') matrix admits the skew-diagonal form
\begin{equation}
  J=\begin{pmatrix}
    0 & J_1 & & & & \\
    -J_1& 0& & & & \\
     & & 0& J_2& & \\
     & &-J_2&0&& \\
     &&&&&\ddots
\end{pmatrix} \ .
\end{equation}

Finally, the functions $F$ and $\Pi$ are given by
\begin{equation}
F=1-\sum_{k=1}^n\frac{a_k^2\mu_k^2}{r^2+a_k^2}\  , \qquad \Pi=\prod_{k=1}^n(r^2+a_k^2) \ ,
\end{equation}
while in order to account for the rotation the $(x_k,y_k)$ coordinates are  written in terms of poly-spherical coordinates as 
\begin{equation}\label{eq:cartesianDodd}
    x_k=\sqrt{r^2 +a_k^2}\mu_k\cos{\phi_k}\ , \qquad y_k=\sqrt{r^2 + a_k^2}\mu_k\sin{\phi_k}\ .
\end{equation}
The condition $\sum_{k=1}^n\mu_k^2=1$  forces the radial variable $r$ to satisfy
\begin{equation}
\sum_{k=1}^n\frac{x_k^2+y_k^2}{r^2+a_k^2}=1 \ .
\end{equation}
Finally with the help of the functions $F$ and $\Pi$ one can write down the gravitational potential as
\begin{equation}\label{eq:potentialDodd}
    \Phi_G=-\frac{\widehat{M} r^2}{\Pi F}  \ . 
\end{equation}

\subsection{MP BH in $D=2n+2$}
In even spacetime dimensions $D=2n+2$, one has to introduce an additional variable $\mu_0$ such that 
\begin{equation}
\sum_{k=1}^n\mu_k^2+\mu_0^2=1\ .
\end{equation}
The MP metric  can be written as
\begin{multline}
    ds^2=dt^2-\frac{\widehat{M} r}{\Pi F}\Bigg(dt+\sum_{k=1}^na_k\mu_k^2d\phi_k\Bigg)^2-\frac{\Pi F}{\Pi-\widehat{M} r}dr^2 
    -\sum_{k=1}^n(r^2+a_k^2)(d\mu_k^2+\mu_k^2d\phi_k^2)-r^2d{\mu_0}^2,
\end{multline}
where now the angular-momentum matrix can be written as
\begin{equation}
  J=\begin{pmatrix}
    0 & J_1 & & & & &\\
    -J_1& 0& & & & &\\
     & & 0& J_2& & &\\
     & &-J_2&0&& &\\
     &&&&&\ddots &\\
     &&&&& & 0
\end{pmatrix}\ ,
\end{equation}
with vanishing last row and column, since $SO(2n+1)$ has rank $n$ as $SO(2n)$, and the $(x_k,y_k)$ and $z$ coordinates are  written in terms of poly-spherical coordinates as  
\begin{equation}
    x_k=\sqrt{r^2 +a_k^2}\mu_k\cos{\phi_k}\ , \qquad y_k=\sqrt{r^2 + a_k^2}\mu_k\sin{\phi_k}\ , \qquad z = r \mu_0\ .
\end{equation}
The condition $\sum_{k=1}^n\mu_k^2=1-\mu_0^2$ forces the radial variable $r$ to satisfy
\begin{equation}
    \frac{z^2}{r^2}+\sum_{k=1}^n\frac{x_k^2+y_k^2}{r^2+a_k^2}=1 \ .
\end{equation}
In this case the gravitational potential has the form
\begin{equation}\label{eq:potentialDeven}
\Phi_G=
    -\frac{\widehat{M} r}{\Pi F}  \ .
\end{equation}

One can easily check that for $n=1$ ($D=4$), one gets the Kerr metric with $a_1=a$, $\mu_1=\sin{\theta}$ and ${\mu_0}=\cos{\theta}$. In both $D$ even and $D$ odd cases, one can check that setting $a_k=0$ one gets the non-rotating Schwarzschild-Tangherlini (ST) BH metric, or setting $\widehat{M}=0$ one gets the flat metric in $D=d+1$ dimensions.

\subsection{MP in the Kerr-Schild gauge}
As for Kerr(-Newman) BHs, the MP solutions in any $D$ can be put in the KS gauge \eqref{eq:forma della metrica di {KS}}, with $\Phi_G$ given in \eqref{eq:potentialDodd} in odd dimensions and in \eqref{eq:potentialDeven} in even dimensions respectively. The null vectors $K_\mu$ are
\begin{equation}
\label{eq:k_muDodd}
    K^{(o)}_\mu=\Big(1,\frac{rx_1+a_1y_1}{r^2+a_1^2},\frac{ry_1-a_1x_1}{r^2+a_1^2},...,\frac{rx_n+a_n y_n}{r^2+a_n^2},\frac{ry_n-a_nx_n}{r^2+a_n^2}\Big)
\end{equation}
for $D=2n+1$ and 
\begin{equation}
\label{eq:k_muDeven}
    K^{(e)}_\mu=\Big(1,\frac{rx_1+a_1y_1}{r^2+a_1^2},\frac{ry_1-a_1x_1}{r^2+a_1^2},...,\frac{rx_n+a_n y_n}{r^2+a_n^2},\frac{ry_n-a_nx_n}{r^2+a_n^2},\frac{z}{r}\Big)
\end{equation}
for $D=2n+2$.
In all cases, the transformation to the KS gauge involves a change of coordinates which does not modify the definition of the radial variable~\cite{Monteiro:2014cda}.  

\section{Cross section for massive scalar on MP BHs in arbitrary dimension D}\label{sec:CrossSection}

In order to probe MP BHs and expose their multi-polar structure we consider the scattering of a massive scalar particle with $m\ll M$. As for Kerr(-Newman) BHs, exploiting the virtues of the KS gauge, the tree-level amplitude is simply given by  
\begin{equation}
\label{eq:definizione ampiezza di diffusione MPBH D}
    2\pi\delta(q_0)i\mathcal{M}(p,p^\prime,\vec{q}\,)={i}\widetilde{h}^{\mu\nu}(q)p_\mu p'_\nu\ ,
\end{equation}
and requires the computation of the FTs 
\begin{equation}
    \widetilde{h}_{\mu\nu}(q)=2\pi\delta(q_0)\int d^{d}xh_{\mu\nu}(\vec{x})e^{i\vec{q}\cdot\vec{x}} \ ,
\end{equation}
where the $2\pi\delta(q_0)$ factor is due to time independence of $h_{\mu\nu}$.
Switching from the coordinates $(x_i, y_i)$, and $z$ in the $D$ even case, to the poly-spherical coordinates $r, \mu_i, \phi_i$ introduced above, one needs the Jacobian, explicitly computed in Appendix~\ref{app:Jacobian}.  Let us discuss the two cases separately.

\subsection{${D_o}=2n+1=d+1$ odd case}
The Jacobian for the change in the measure is given by\footnote{In our conventions we express $\mu_n$ in terms of the other $\mu_k$. This explains why it does not appear in the Jacobian and in the measure.}
\begin{equation}
    {\cal J}_{o}=\frac{\Pi F}{r}\prod_{k\neq n}^{\frac{D-1}{2}}\mu_k\ ,
\end{equation}
so that 
\begin{equation}
    \widetilde{h}_{\mu\nu}(\vec{q}\,)=\int \frac{\Pi F}{r}dr\Big(\prod_{k\neq n}^{\frac{D-1}{2}}\mu_kd\mu_k\Big)\Big(\prod_{l=1}^{\frac{D-1}{2}}d\phi_l\Big)\frac{\widehat{M} r^2}{\Pi F}K_\mu K_\nu e^{-i\vec{q}\cdot\vec{x}} \ .
\end{equation}
As in the case of Kerr BHs, we  find it convenient to write the exponent as ${\vec{q}\cdot\vec{x} = \vec{u}\cdot\vec{n} = u \cos\tilde\theta}$, where $\vec{n}$ is a unit vector and 
\begin{equation}
\vec{u} = (q_{x_1} \sqrt{r^2+a_1^2}, q_{y_1} \sqrt{r^2+a_1^2}, ..., q_{x_n} \sqrt{r^2+a_n^2}, q_{y_n} \sqrt{r^2+a_n^2} ) \ .
\end{equation}
Moreover, we define the vector $\vec{u}_a$ with components
\begin{equation}
    u^i_a =a^i{}_j q^j\ , 
\end{equation}
where the matrix $a=J/M$ is defined in \eqref{eq:DefSpinParam} and so that $u^2=|\vec{u}|^2={r^2|\vec{q}|^2+u_a^2}$, where $|\vec{q}|^2 = \Sigma_{k=1}^n |\vec{q}_{(k)}|^2$ with  $ |\vec{q}_{(k)}|^2 = {q}_{(k)}^2 = q_{x_k}^2+q_{y_k}^2$. 
Making $\widehat{M}$ explicit one finds
\begin{equation}
\label{eq:passaggio intermedio angolare calcolo hmunu D dim}
    \widetilde{h}_{\mu\nu}(\vec{q}\,)=-\frac{16\pi GM}{D-2}\frac{\Omega_{D-3}}{\Omega_{D-2}}\int rdrd\tilde{\theta}\sin^{D-3}{\tilde{\theta}}K_\mu K_\nu e^{-iu\cos{\tilde{\theta}}} \ ,
\end{equation}
 where $\Omega_{D-3}$ arises from the `trivial' integration over the angular variables on the $(D-3)$-sphere `orthogonal' to the polar angle $\tilde\theta$. The $\tilde{\theta}$ integral can then be performed iteratively by parts. Taking into account the vanishing of the boundary terms one finds
\begin{multline}
     {\cal I}_{D{-}3}(u) = \int_0^{\pi} {d\tilde{\theta}} \sin^{D-3}{\tilde{\theta}} e^{-iu\cos{\tilde{\theta}}}
  =\frac{D-4}{u}\int_0^\pi d\tilde{\theta}\sin^{D-5}{\tilde{\theta}}(-i\cos{\tilde{\theta}})e^{-iu\cos{\tilde{\theta}}} \ ,
\end{multline}
that can be written in the form
\begin{equation}
{\cal I}_{D{-}3}(u) = \frac{D-4}{u}\int_0^\pi d\tilde{\theta}\sin^{D-5}{\tilde{\theta}}\frac{d}{du}\Big(e^{-iu\cos{\tilde{\theta}}}\Big) = (D-4)\frac{d {\cal I}_{D{-}5}(u)}{udu} \ .
\end{equation}
Proceeding this way $(D-5)/2 = n-2$ times one finally has 
\begin{equation}
    {\cal I}_{D{-}3}(u) = \pi(D-4)!!\Big(\frac{1}{u}\frac{d}{du}\Big)^{\frac{D-3}{2}}J_0(u)\ .
\end{equation}
Before plugging into the FT it is expedient to use the recursion relation
\begin{equation}
\label{eq:proprietà ricorsiva derivate bessel}
    \Big(\frac{1}{u}\frac{d}{du}\Big)^k(u^{-{\alpha}}J_{\alpha}(u))=(-1)^k u^{-{\alpha}-k}J_{{\alpha}+k}(u)
\end{equation}
and get  
\begin{equation}
    \widetilde{h}_{\mu\nu}(\vec{q})=-\frac{16\pi^2GM}{D-2}\frac{\Omega_{D-3}}{\Omega_{D-2}}(D-4)!!\int_0^\infty rdr K_\mu K_\nu\frac{J_{\tfrac{D-3}{2}}(u)}{u^{\tfrac{D-3}{2}}}\ .
\end{equation}

\subsection{$D_e=2n+2=d+1$ even case}

Taking into account the extra coordinate $z=r{\mu_0}$ and the modified Jacobian
\begin{equation}
 {\cal J}_{e}=\Bigg(\prod_{k\neq n}^{\frac{D-2}{2}}\mu_k\Bigg) F\Pi
\end{equation} 
for the change to  poly-spherical coordinates, combining volume element and gravitational potential yields
\begin{equation}
\label{eq:passaggio intermedio caso pari}
 \widetilde{h}_{\mu\nu}(\vec{q}\,)=\int\Pi Fdr d{\mu_0}\Big(\prod_{k\neq n}^{\frac{D-2}{2}}\mu_kd\mu_k\Big)\Big(\prod_{l=1}^{\frac{D-2}{2}}d\phi_l\Big)\frac{\widehat{M} r}{\Pi F}K_\mu K_\nu e^{-i\vec{q}\cdot\vec{x}}\ .
\end{equation}
Simplifying and changing angular coordinates to poly-spherical ones yield the same integral as \eqref{eq:passaggio intermedio angolare calcolo hmunu D dim}. Following the same  procedure as in the odd $D$ case, one integrates over $\tilde{\theta}$, then after $\frac{D-4}{2}=n-1$ integrations by part, the angular integral reads
\begin{equation}
    (D-4)!!\Bigg(\frac{1}{u}\frac{d}{du}\Bigg)^{\frac{D-4}{2}}\int_0^\pi \sin{\tilde{\theta}}e^{-iu\cos{\tilde{\theta}}}d\tilde{\theta} 
    = (D-4)!!2 \Bigg(\frac{1}{u}\frac{d}{du}\Bigg)^{\frac{D-4}{2}} j_0(u)\ .
\end{equation}
Recalling that $j_\ell(u)=\sqrt{\frac{\pi}{2u}}J_{\ell+\frac{1}{2}}(u)$, and perusing \eqref{eq:proprietà ricorsiva derivate bessel} allows one to rewrite the integral  \eqref{eq:passaggio intermedio caso pari} as
\begin{align}
    \widetilde{h}_{\mu\nu}(\vec{q}\,)&=-\frac{16\pi GM}{D-2}\frac{\Omega_{D-3}}{\Omega_{D-2}}(D-4)!!\int_0^\infty rdr\sqrt{2\pi}K_\mu K_\nu\frac{J_{\tfrac{D-3}{2}}(u)}{u^{\tfrac{D-3}{2}}}
    \nonumber \\
   &=-\frac{32\pi GM}{D-2}\frac{\Omega_{D-3}}{\Omega_{D-2}}(D-4)!!\int_0^\infty rdrK_\mu K_\nu\frac{j_{\tfrac{D-4}{2}}(u)}{u^{\tfrac{D-4}{2}}}\ .
\end{align}
As a `sanity' check, putting $D=4$ one gets $j_0(u)$ for Kerr, as expected. Notice how the FT of the metric is built by spherical Bessel functions in the $D=\mathrm{even}$ case and by Bessel functions in the $D=\mathrm{odd}$ case, as already noticed in~\cite{Bianchi:2024vmi} where the gravitational form factors in higher dimensions were studied.

\subsection{Computing the radial integrals}\label{Sec:RadialIntegral}

Aside from the overall coefficient $C_D$ 
\begin{equation}\label{eq:Coefficiente}
C_D=- GM\pi\frac{2^{\frac{D+5}{2}} \Gamma(\tfrac{D-1}{2})}{D-2} \quad ,
\end{equation}
the integrals for the two cases can be performed in the same way as in
\begin{equation}
\widetilde{h}_{\mu\nu}=C_D\int_0^\infty rdr K_\mu K_\nu \frac{J_{\alpha}(u)}{u^{\alpha}}\ ,
\end{equation}
where for compactness of notation 
\begin{equation}
    \alpha = \frac{D-3}{2}\ .
\end{equation}
While the details of the computation are relegated to Appendix~\ref{app:ExpCalc}, the final result reads
\begin{align}
\widetilde{h}_{00}&=C_D\frac{1}{|\vec{q}\, |^2}\frac{J_{{\alpha}-1}(u_a)}{u_a^{{\alpha}-1}}\ ,\\ 
\widetilde{h}_{0i}&=C_D\Big[-\frac{iq_i}{|\vec{q}\, |^3}\sqrt{\frac{\pi}{2}}\,\frac{J_{{\alpha}-\frac{1}{2}}(u_a)}{u_a^{{\alpha}-\frac{1}{2}}}+\frac{ia_{ik}q^k}{|\vec{q}\, |^2}\frac{J_{\alpha}(u_a)}{u_a^{\alpha}}\Big]\ ,\\
\widetilde{h}_{ij}&=C_D\left[\frac{\delta_{ij}}{|\vec{q}\, |^2}\frac{J_{\alpha}(u_a)}{u_a^{\alpha}}-\frac{2q_iq_j}{|\vec{q}\, |^4}\frac{J_{\alpha}(u_a)}{u_a^{\alpha}}-\frac{a_{ik}q^ka_{jk}q^k}{|\vec{q}\, |^2}\frac{J_{\alpha+1}(u_a)}{u_a^{{\alpha}+1}}\right.\\&\nonumber \left. +(a_{ik}q^kq_j+a_{jk}q^kq_i)\frac{1}{|\vec{q}\, |^3}\sqrt{\frac{\pi}{2}}\frac{J_{{\alpha}+\frac{1}{2}}(u_a)}{u_a^{{\alpha}+\frac{1}{2}}}\right]\ .
\end{align}
The `off-shell' amplitude (with $E=E'$ since the process is elastic) is then
\begin{align}
&i {\cal M} = i {C_D \over |\vec{q}\, |^2}\Bigg\{E^2\frac{J_{{\alpha}{-}1}(u_a)}{u_a^{{\alpha}{-}1}} 
-iE\left[\frac{\vec{q}{\cdot}(\vec{p}{+}\vec{p}\, ')}{|\vec{q}\, |}\sqrt{\frac{\pi}{2}}\frac{J_{{\alpha}{-}\frac{1}{2}}(u_a)}{u_a^{{\alpha}{-}\frac{1}{2}}}-{\vec{u}_a{\cdot}(\vec{p} {+}\vec{p}\, ')}\frac{J_{\alpha}(u_a)}{u_a^{\alpha}}\right]\nonumber \\
&\quad +\Big[\left({\vec{p}{\cdot}\vec{p}\, '}{-}2 \frac{\vec{q}{\cdot}\vec{p} \vec{q}{\cdot}\vec{p}\, '}{|\vec{q}\, |^2}\right)\frac{J_{\alpha}(u_a)}{u_a^{\alpha}} -{\vec{u}_a{\cdot}\vec{p} \vec{u}_a{\cdot}\vec{p}\, '}\frac{J_{\alpha+1}(u_a)}{u_a^{{\alpha}+1}} +\frac{\vec{u}_a{\cdot}\vec{p} \vec{q}{\cdot}\vec{p}\, '{+}\vec{u}_a{\cdot}\vec{p}\, ' \vec{q}{\cdot}\vec{p}}{|\vec{q}\, |^3}\sqrt{\frac{\pi}{2}}\frac{J_{{\alpha}+\frac{1}{2}}(u_a)}{u_a^{{\alpha}+\frac{1}{2}}}\Big] 
\Bigg\} \ .
\end{align}
Going on-shell, $|\vec{p}| = |\vec{p}\, '| $, so that $\vec{q}\cdot (\vec{p}+\vec{p}\, ')=0$, and all the Bessel functions of order $\alpha \pm {1\over 2}$ disappear and one gets
\begin{equation}
\begin{aligned}
i {\cal M}_{on-shell} &= i {C_D \over |\vec{q}|^2}\Bigg\{E^2 \frac{J_{{\alpha}{-}1}(u_a)}{u_a^{{\alpha}{-}1}} 
+i E {\vec{u}_a{\cdot}(\vec{p}{+}\vec{p}\, ')}\frac{J_{\alpha}(u_a)}{u_a^{\alpha}} \\
&+\Big[\left({\vec{p}{\cdot}\vec{p}\, '}{-}2 \frac{\vec{q}{\cdot}\vec{p} \vec{q}{\cdot}\vec{p}\, '}{|\vec{q}|^2}\right)\frac{J_{\alpha}(u_a)}{u_a^{\alpha}} -{\vec{u}_a{\cdot}\vec{p} \vec{u}_a{\cdot}\vec{p}\, '}\frac{J_{\alpha+1}(u_a)}{u_a^{{\alpha}+1}} \Big] 
\Bigg\} \ .
\end{aligned}
\end{equation}
Applying the Fermi golden rule one finds the differential cross-section
\begin{equation}
\begin{aligned}
    \frac{d\sigma}{d\Omega_{D-2}}&= \frac{2\pi}{2Ev}|\mathcal{M}_{on-shell}|^2\delta(E-E^\prime)\frac{|\vec{p^\prime}\,|^{D-2}d|\vec{p^\prime}\,|}{(2\pi)^{D-1}2E^\prime}\\
    & =\frac{|\vec{p}\,|^{D-4}}{2(2\pi)^{D-2}}\frac{C_D^2}{|\vec{q}|^4}\Bigg|E^2 \frac{J_{{\alpha}-1}(u_a)}{u_a^{{\alpha}-1}}
+iE\Big({\vec{u}_a}\cdot {\vec{P}}\,\Big)\frac{J_{\alpha}(u_a)}{u_a^{\alpha}} \\
& +\left[\vec{p}\cdot\vec{p^\prime}-2\frac{(\vec{q}\cdot\vec{p}\,)(\vec{q}\cdot\vec{p^\prime}\,)}{|\vec{q}|^2}\right]\frac{J_{\alpha}(u_a)}{u_a^{\alpha}}+({\vec{u}_a}\cdot\vec{p}\,)^2\frac{J_{{\alpha}+1}(u_a)}{u_a^{{\alpha}+1}}\Bigg|^2.
\end{aligned}
\end{equation}

Using then $|\vec{p}\,|=Ev$  and integrating over $E^\prime$ yields
\begin{equation}
    \frac{d\sigma}{d\Omega_{D-2}}=\frac{|\vec{p}\,|^{D-4}}{4(2\pi)^{D-2}}|\mathcal{M}_{on-shell}|^2 \ ,
\end{equation}
and one arrives at the final compact Rutherford-like formula 
\begin{equation}
\begin{aligned}
\frac{d\sigma}{d\Omega_{D-2}}&= \frac{|\vec{p}\,|^{D-4}}{4(2\pi)^{D-2}}\frac{C_D^2}{|\vec{q}|^4}\Bigg|E^2 \frac{J_{{\alpha}-1}(u_a)}{u_a^{{\alpha}-1}}
+iE\Big({\vec{u}_a}\cdot {\vec{P}}\,\Big)\frac{J_{\alpha}(u_a)}{u_a^{\alpha}} \\
& +\left(\vec{p}\cdot\vec{p^\prime}-\frac{2(\vec{q}\cdot\vec{p}\,)(\vec{q}\cdot\vec{p^\prime}\,)}{|\vec{q}|^2}\right)\frac{J_{\alpha}(u_a)}{u_a^{\alpha}}+({\vec{u}_a}\cdot\vec{p}\,)^2\frac{J_{{\alpha}+1}(u_a)}{u_a^{{\alpha}+1}}\Bigg|^2 \ .
\end{aligned}
\end{equation}
As in the $D=4$ Kerr case one can analyze the effect of the BH angular momentum coded in the vector $\vec{u}_a$. For simplicity let us consider small deflection angles, corresponding to the eikonal limit with large impact parameters and small $|\vec{q}|$. Choosing $\vec{p}$ and $\vec{p}\, ^\prime$ in the $(x_i,y_i)$ plane with $\vec{q}=\vec{p}-\vec{p^\prime}$, along $x_i$ and  ${\vec{P}}= \vec{p}+\vec{p}\, '$ along $y_i$ one has
\begin{equation}
    |\vec{q}\,|=2|\vec{p}\,|\sin{\frac{\vartheta}{2}} \qquad |{\vec{P}}\,|=2|\vec{p}\,|\cos{\frac{\vartheta}{2}}\ .
\end{equation}
Recalling that $\vec{p}\sim {\vec{P}}/2$ (for small $\vartheta$) and $E v=|\vec{p}\,|$ one finds
\begin{equation}
\begin{aligned}
    \frac{d\sigma}{d\Omega_{D-2}}=& \frac{E^{D-4}v^{D-8}}{64(2\pi)^{D-2}}\frac{C_D^2}{\sin^4{\frac{\vartheta}{2}}}\Bigg|\frac{J_{{\alpha}-1}(u_a)}{u_a^{{\alpha}-1}}
+i\Big({\vec{u}_a}\cdot \vec{v}\,\Big)\frac{J_{\alpha}(u_a)}{u_a^{\alpha}} \\
& +v^2\frac{J_{\alpha}(u_a)}{u_a^{\alpha}}-\frac{2(\vec{q}\cdot\vec{v}\,)}{|\vec{q}|^2}\frac{J_{\alpha}(u_a)}{u_a^{\alpha}}-({\vec{u}_a}\cdot\vec{v}\,)^2\frac{J_{{\alpha}+1}(u_a)}{u_a^{{\alpha}+1}}\Bigg|^2\ .
\end{aligned}
\end{equation}
First of all the strict limit $a_i\rightarrow 0$ produces the gravitational Rutherford formula for Schwarzschild-Tangherlini BHs in $D$ dimensions, {\it viz.}
\begin{equation}
\frac{d\sigma}{d\Omega_{D-2}}\Bigg|_{a_i=0}=\left\{\frac{C_D^2}{64(2\pi)^{D-2}}\frac{1}{2^{2\alpha}\Gamma(\alpha+1)^2} \right\}\frac{E^{D-4}v^{D-8}}{\sin^4{\frac{\vartheta}{2}}} 
\Bigg|2\alpha +\Bigg[v^2-\frac{2(\vec{v}\cdot\vec{q}\,)^2}{|\vec{q}|^2}\Bigg]\Bigg|^2\ ,
\end{equation}
where the overall coefficient simplifies to
\begin{equation}
    \frac{C_D^2}{64(2\pi)^{D-2} 2^{2\alpha}\Gamma(\alpha+1)^2} = \frac{ G^2M^2}{(2\pi)^{D-4}(D-2)^2} \ .
\end{equation}

Expanding to second order in $u_a$ the amplitude yields
\begin{equation}
\begin{aligned}
    &\frac{d\sigma}{d\Omega_{D-2}}= \frac{ G^2M^2}{(2\pi)^{D-4}(D-2)^2} \frac{ E^{D-4}v^{D-8}}{\sin^4{\frac{\vartheta}{2}}} \Bigg|\Bigg(D-3-{u_a^2}\Bigg) \\
    &+\Bigg[v^2-\frac{2(\vec{v}\cdot\vec{q}\,)^2}{|\vec{q}|^2}\Bigg]\Bigg(1 -\frac{u_a^2}{D-1}\Bigg) +i({\vec{u}_a}\cdot\vec{v}\,)
    -({\vec{u}_a}\cdot \vec{v}\,)^2\Bigg(\frac{1}{D-1}\Bigg)+\mathcal{O}(u_a^2)\Bigg|^2 \ .
\end{aligned}
\end{equation}
The (second order) correction to the cross-section comes from the square of the purely imaginary term $\sim i {\vec{u}_a}\cdot\vec{v}$ and from the beating term between the real terms of order zero in $u_a$ and those of order $u_a^2$. 

\section{Multipoles in higher dimensions: stress multipoles}\label{sec:MultiRev}
In order to relate the information coded in the scattering amplitude to the properties of the source, let us review the analysis of the multipolar structure of higher-dimensional compact objects performed in~\cite{Bianchi:2024shc}. The main result is that a new class of multipoles, dubbed `stress multipoles', appear that together with the known mass and current multipoles~\cite{Geroch:1970cc, Geroch:1970cd, Hansen:1974zz, Thorne:1980ru} characterize the asymptotic behavior of stationary asymptotically flat metrics\footnote{Note that in this paper we are using opposite signature compared to~\cite{Bianchi:2024shc}.} as 
\begin{equation}\label{eq:MultipoleExpandedMetric}
\begin{aligned}
    g_{00}&=1-4\frac{d-2}{d-1}\sum_{\ell=0}^{+\infty}\frac{G M \rho(r)}{r^\ell}\mathbb{M}^{(\ell)}_{A_\ell}N_{A_\ell}+\cdots\ ,\\
    g_{0i}&=-2(d-2)\sum_{\ell=1}^{+\infty}\frac{GM\rho(r)}{r^\ell}\mathbb{J}^{(\ell)}_{i, A_\ell}N_{A_\ell}+\cdots\ ,\\ 
    g_{ij}&=-\delta_{ij}-4\frac{d-2}{d-1}\sum_{\ell=2}^{+\infty}\frac{G M \rho(r)}{r^\ell}\tilde{\mathbb{G}}^{(\ell)}_{ij,A_\ell}N_{A_\ell}+\cdots\ ,
\end{aligned}
\end{equation}
where $\ell$ organizes the multipolar expansion, ${\rho(r) = \Gamma(\alpha)/\pi^\alpha r^{2\alpha}}$, and, following the standard notation, repeated indices are denoted by   $A_\ell=a_1\cdots a_\ell$,  $N_{A_\ell}=\frac{x_{a_1}\cdots \,x_{a_\ell}}{r^\ell}$,  while $\mathbb{M}^{(\ell)}_{A_\ell}$ and $\mathbb{J}^{(\ell)}_{i, A_\ell}$ are the mass and current multipole moments, respectively. The additional tensor $\tilde{\mathbb{G}}^{(\ell)}_{ij,A_\ell}$ is related to the stress multipoles through  
\begin{equation}
\mathbb{G}^{(\ell)}_{ij,A_\ell}=\tilde{\mathbb{G}}^{(\ell)}_{ij,A_\ell}+\frac{1}{2}\delta_{ij}\Big(\mathbb{M}^{(\ell)}_{A_\ell}-\tilde{\mathbb{G}}^{(\ell)}_{kk,A_\ell}\Big)\ ,
\end{equation}
which vanish in $D=4$. For those compact objects whose multipolar structure is uniquely determined by the mass $M$ and the angular momentum $J$ (spin-induced multipole moments), such multipoles can be written in terms of three classes of `form factors' $F_{2\ell , 1}$, $F_{2\ell,2}$ and $F_{2\ell +1,3}$ as
\begin{equation}\label{eq:GravitationalMultipoles}
    \begin{aligned}
\mathbb{M}^{(2\ell)}_{A_{2\ell}}&=\frac{(d+4\ell-4)!!}{(d-2)!!}(-1)^\ell\Big(F_{2\ell, 2}+(d-2)F_{2\ell, 1}\Big)(-S\cdot S)_{A_{2\ell}}\Big|_{\rm STF}\ ,\\
\mathbb{J}^{(2\ell+1)}_{i,A_{2\ell+1}}&=\frac{(d+4\ell-2)!!}{(d-2)!!}(-1)^\ell F_{2\ell+1, 3} \ S_{ia_1}(-S\cdot S)_{A_{2\ell}}|_{\rm ASTF}\ , \\
\mathbb{G}^{(2\ell)}_{ij,A_{2\ell}}&=(d-1)\frac{(d+4\ell-4)!!}{(d-2)!!}(-1)^\ell F_{2\ell, 2} \ S_{ia_1}S_{ja_2}(-S\cdot S)_{A_{2\ell-2}}|_{\rm RSTF}\ .
    \end{aligned}
\end{equation}
where $S=J/M$ is the spin-density matrix and STF means totally symmetric trace-free, while ASTF and RSTF denote irreducible representations of $SO(d)$ with one or two antisymmetrizations~\cite{Heynen:2023sin,Gambino:2024uge}. In particular, stress multipoles are RSTF (Riemann-symmetric and trace-free), namely STF in the $A_{2\ell-2}$, with the same symmetries as the Riemann tensor for all $(ia_1,ja_2)$ (including Bianchi identities), traceless with respect to all indices and with zero anti-symmetrization of any three indices.
Such form factors characterize the stress tensor of the source in momentum space.

Here we are interested in the multipolar structure of MP BHs, whose linearized source in arbitrary dimension gives the constant form factors
\begin{equation}\label{eq:FFMParbitraryD}
\begin{aligned}
    F_{2\ell+2, 2}&=-\frac{1}{2}\Bigg(\frac{\Gamma(d/2)}{2^{2-d}(d-1)^{\frac{d-2}{2}}}\Bigg) \frac{(-1)^\ell}{\ell!\ \Gamma\left(\ell+2+\frac{d-2}{2}\right)}\left(\frac{d-1}{4}\right)^{2\ell+1+\frac{d-2}{2}}\ , \\
    F_{2\ell+1, 3}&=\Bigg(\frac{\Gamma(d/2)}{2^{2-d}(d-1)^{\frac{d-2}{2}}}\Bigg) \frac{(-1)^\ell}{\ell!\ \Gamma\left(\ell+1+\frac{d-2}{2}\right)}\left(\frac{d-1}{4}\right)^{2\ell+\frac{d-2}{2}}\ ,\\
    F_{2\ell+2, 1}&=F_{2\ell+2, 2}+F_{2\ell+1, 3}\ .
\end{aligned}
\end{equation}
Plugging this into eq. \eqref{eq:GravitationalMultipoles} one obtains 
\begin{equation}
    \begin{aligned}
        \mathbb{M}^{(2\ell+2)}_{A_{2\ell+2}}&=\frac{d-1}{4}\frac{(d+4\ell)!!\ (d+2\ell)}{(d-2)!!\ (\ell+1)!\ \Gamma\left(\ell+2+\frac{d-2}{2}\right)}\left(\frac{d-1}{4}\right)^{2\ell+1+\frac{d-2}{2}}(-S\cdot S)_{A_{2\ell+2}}\Big|_{\rm STF}\ , \\
         \mathbb{J}^{(2\ell+1)}_{i,A_{2\ell+1}}&=\frac{(d+4\ell-2)!!}{(d-2)!!\ \ell!\ \Gamma\left(\ell+1+\frac{d-2}{2}\right)}\left(\frac{d-1}{4}\right)^{2\ell+\frac{d-2}{2}} \ S_{ia_1}(-S\cdot S)_{A_{2\ell}}|_{\rm ASTF}\ , \\
         \mathbb{G}^{(2\ell+2)}_{ij,A_{2\ell+2}}&=\frac{d-1}{2}\frac{(d+4\ell)!!}{(d-2)!!\ \ell!\ \Gamma\left(\ell+2+\frac{d-2}{2}\right)}\left(\frac{d-1}{4}\right)^{2\ell+1+\frac{d-2}{2}} \ S_{ia_1}S_{ja_2}(-S\cdot S)_{A_{2\ell}}|_{\rm RSTF}\ ,
    \end{aligned}
\end{equation}
where the acronym STF, ASTF and RSTF are defined above and denote irreducible representations of $SO(d)$~\cite{Heynen:2023sin, Gambino:2024uge}.
For later comparison with scattering amplitudes, it is worth noticing that the form factors in eq. \eqref{eq:FFMParbitraryD} can be packaged into generating functions
\begin{equation}
\begin{aligned}
    F_2^{(d)}(\zeta)&=-\frac{1}{2}\zeta\mathcal{Z}_1^{(d)}(\zeta)\ ,\\
    F_3^{(d)}(\zeta)&= \mathcal{Z}_0^{(d)}\ ,\\
    F_1^{(d)}(\zeta)&=F_2^{(d)}(\zeta)+F_3^{(d)}(\zeta)\ .
\end{aligned}
\end{equation}
These functions can be expressed in terms of Bessel functions
\begin{equation}
    \mathcal{Z}_n^{(d)}(\zeta)     
    =\Bigg(\frac{\Gamma(d/2)}{2^{2-d}(d-1)^{\frac{d-2}{2}}}\Bigg) \zeta^{-\frac{d-2}{2}}J_{n+\frac{d-2}{2}}\left(\frac{d-1}{2}\zeta\right)\ .
\end{equation}
where for comparison with the scattering amplitudes one has to identify $\zeta=2 u_a/ (d-1)$.
Our aim then is to investigate the scattering of a massive scalar probe off a MP BH. Later on we will focus on the five-dimensional case,  whereby there are two angular momenta. The mass, the current and stress multipoles are proportional to
\begin{equation}
    \mathbb{M}_{2\ell}\propto(a_1^2-a_2^2)^{2\ell}\ ,\quad \mathbb{J}_{2\ell+1}\propto a_i(a_1^2-a_2^2)^{2\ell}\ ,\quad \mathbb{G}_{2\ell+2}\propto a_1a_2(a_1^2-a_2^2)^{2\ell}
\end{equation}
respectively.  To have a clear understanding of how the presence of the different multipoles affect the scattering process, we will analyze two extreme limits, namely one in which the two angular momenta are equal or opposite, and one in which one of the angular momenta (say $a_2$) vanishes. These configurations allow for analytical and compact expression in order to focus on the physics.  
The former case, namely $a_1=\pm a_2=a$, is particularly simple and interesting because the metric becomes cohomogeneity-one (with full rotational symmetry, despite the non vanishing angular momenta!) and all the gravitational multipoles vanish except for the current dipole and the stress quadrupole. Therefore this limit is the simplest setup for studying the effect of stress multipoles on a physical observable such as the eikonal phase. From the comparison of the result with the $D=4$ counter-part it is possible to elucidate why stress multipoles are not physical in $D=4$. On the other hand, in the latter case the stress multipoles vanish, and thus this limit is the one that resembles Kerr BHs the most, as will be appreciated from the computation of the eikonal phase.

Finally, in order to streamline the physical meaning of the gravitational form factors defined in \eqref{eq:GravitationalMultipoles}, it is amusing to make a comparison with the standard definition of form factors in elastic electron-proton scattering 
\begin{equation}
{\cal M}^{\rm extended}_{fi} = {\cal M}^{\rm point}_{fi} F(\vec{q})\ ,
\end{equation}
with  
\begin{equation}
F(\vec{q}) = {\int d^3 x e^{-i\vec{q}\cdot \vec{x}} \varrho(\vec{x})\over \int d^3 x \varrho(\vec{x})} \ ,
\end{equation}
where $\varrho(\vec{x})$ is the static proton `charge' density, so that the proton form factor $F(\vec q\, )$ is normalized to $F(0) = 1$. Including the recoil of the proton ($E_e\neq E_e'$) and, possibly, other  inelastic effects,  one should take into account that $q^2\neq -|\vec{q}\, |^2$, {\it viz.}
\begin{equation}
-|\vec{q}|^2 = q^2 \left[ 1 - {q^2\over 4M_p^2} \right] = q^2 - (E_e-E_e')^2 \ .
\end{equation}
As a result Rosenbluth formula for inelastic e-p scattering is usually written in terms of `structure functions' $G_E(q^2)$ and $G_M(q^2)$ with $G_E(q^2)\approx F(\vec{q}\, )$ for $|\vec{q}\, |^2\ll M_p^2$, while $G_M(q^2)$ encodes the magnetic moment distribution inside the proton. Notice that while $G_E(0) =1= F(0)$,  $G_M(0) = g_p \approx 2.79 \neq 2$ that shows that protons are not an elementary spin 1/2 particles, whereby $g-2$ would be very small. Quite remarkably the gyromagnetic factor of a Kerr-Newman BH is exactly 2, meaning that BHs are fundamental point-like objects in GR. 

In order to elucidate this statement, fixing the gravitational form factors as in Eq. \eqref{eq:FFMParbitraryD} in order to describe a Kerr BH, we can promote the constant gravitational form factors to having a non trivial local behavior as
\begin{equation}
    F_{i}(aq_\perp)\rightarrow F_{i}(aq_\perp) K(\vec q\, ^2)\ ,
\end{equation}
where $i=1, 2, 3$ and $K(\vec q\, ^2)$ is related to the energy density of the stationary source through 
\begin{equation}
    K(\vec q\, ^2)=\frac{\int d^3x e^{-i\vec q\cdot \vec x} \epsilon(\vec x^2)}{\int d^3x\  \epsilon(\vec x^2)}\ ,
\end{equation}
and is normalized to $K(0)=1$.

So then, just as in particle physics, starting from BHs, which are point-like objects, we can generalize scattering amplitudes involving BHs to processes involving ECOs or BHs mimickers just by considering 
\begin{equation}
{\cal M}^{\rm mimicker}_{fi} ={\cal M}^{\rm Kerr}_{fi} K(\vec q\, ^2)\ .
\end{equation}
An explicit example of gravitational structure function that sources a Kerr mimicker is given in~\cite{Gambino:2025xdr}, in which is studied the case
\begin{equation}\label{eq:KMimicker}
    K(\vec q\, ^2)=e^{-R^2\vec q\, ^2}\ ,
\end{equation}
where $R$ is a new fundamental typical length-scale of the system. It is possible to show that such gaussian smearing of the BH singularity leads to a physically reasonable horizonless Kerr mimicker with gaussian-like behavior, differing from a BH only at the would-be horizon scales, and hence sharing the same multipolar structure~\cite{Gambino:2025xdr}. 

Summarizing, in gravitational processes scattering amplitudes are able to directly probe the BH paradigm by measuring the structure function $K(\vec q\, ^2)$. Indeed, since GR predict BHs to be fundamental point-like objects, structure functions for BHs are exactly $K^{BH}(\vec q\, ^2)=1$, while a non-trivial behavior corresponds to horizonless objects that differ from BH in the would-be horizon region, but resemble BH geometries asymptotically. 
    
\section{A case study: scattering off MP BHs in $D=5$}\label{sec:D5Case}
In order to further investigate the physical meaning of form factors and their relation to the multipolar structure, we now focus on the five-dimensional MP BH and use the results obtained in the previous sections to compute the eikonal phase at leading order in this case.  Through the study of gauge-invariant gravitational observables, such as the eikonal phase, in which form factors explicitly appears in physical quantities, we would like to streamline in particular the role of the stress multipoles for MP BHs that are absent (or not gauge invariant) in lower dimensions {\it e.g.} for Kerr. 

In the case of  $D=5$,  $i.e.$ $n=2$, the null vector, given in general in \eqref{eq:k_muDodd}, becomes
\begin{equation}
\label{eq:k_mu}
    K_\mu=\Big(1,\frac{rx_1+a_1y_1}{r^2+a_1^2},\frac{ry_1-a_1x_1}{r^2+a_1^2},\frac{rx_2+a_2y_2}{r^2+a_2^2},\frac{ry_2-a_2x_2}{r^2+a_2^2}\Big) \ .\end{equation}
Setting $\mu_1=\sin\theta$ and $\mu_2=\cos\theta$, the `cartesian' coordinates in \eqref{eq:cartesianDodd} become 
\begin{equation}
x_1+iy_1=\sqrt{r^2+a_1^2}\sin{\theta}e^{i\phi_1}, \qquad
x_2+iy_2=\sqrt{r^2+a_2^2}\cos{\theta}e^{i\phi_2}.
\end{equation}
and the gravitational potential \eqref{eq:potentialDodd} becomes
\begin{equation}
\Phi_G = -\frac{\widehat{M}}{r^2+a_1^2\cos^2{\theta}+a_2^2\sin^2{\theta}} \ .
\end{equation}
We briefly go through the computation of the FT of the metric that we have performed in detail in Section 4 in arbitrary dimension. The space volume element
\begin{equation}  dx_1dx_2dy_1dy_2=r\sin{\theta}\cos{\theta}(r^2+a_1^2\cos^2{\theta}+a_2^2\sin^2{\theta})drd\theta d\phi_1 d\phi_2.
\end{equation}
nicely combines with the denominator  of $\Phi_G$ to give
\begin{equation}
    \widetilde{h}_{\mu\nu}(\vec{q})=-\frac{16\pi GM}{6\pi}\int rdr\sin{\theta}\cos{\theta}d\theta d\phi_1 d\phi_2 K_\mu K_\nu e^{- i\vec{q}\cdot\vec{x}}.
 \end{equation}
It is convenient to write the exponent as 
\begin{equation}
\vec{q}\cdot\vec{x}=q_{x_1}x_1+q_{x_2}x_2+q_{y_1}y_1+q_{y_2}y_2=\vec{u}\cdot\vec{n},
\end{equation}
where $\vec{n}$ is a unit vector depending on the angular coordinates and 
\begin{equation}
\vec{u}=(q_{x_1}\sqrt{r^2+a_1^2},q_{y_1}\sqrt{r^2+a_1^2}, q_{x_2}\sqrt{r^2+a_2^2},q_{y_2}\sqrt{r^2+a_2^2})
\end{equation}
only depends on the radial coordinate and $\vec{q}$, so that 
\begin{equation}
|\vec{u}\,|^2=|\vec{q}\, |^2(r^2+a_1^2\cos^2{\theta}+a_2^2\sin^2{\theta})=(r^2|\vec{q}\, |^2+a_1^2 q_{{(1)}}^2+a_2^2 q_{{(2)}}^2) .   
\end{equation}
with 
\begin{equation}|\vec{q}\, |^2=(q_{x_1}^2+q_{y_1}^2)+(q_{x_2}^2+q_{y_2}^2)\equiv q_{{(1)}}^2+q_{{(2)}}^2 ,
\end{equation}
and to express the scalar product as 
\begin{equation}
    \vec{u}\cdot\vec{n}= |\vec{u}|\cos{\tilde{\theta}} ,
\end{equation}
where  $\tilde{\theta}$ is the angle between $\vec{u}$ and $\vec{n}$. We  then rewrite the measure on the unit 3-sphere $d\Omega_3$ in the more convenient set of `hyper-spherical' coordinates and get
\begin{equation}
    d\Omega_3=\sin{\theta}\cos{\theta}d\theta d\phi_1d\phi_2 =\sin^2{\tilde{\theta}}d\tilde{\theta}d\Omega_2,
\end{equation}
where $d\Omega_2$ is the measure on the unit 2-sphere that can be immediately integrated to $4\pi$. One can then integrate over the polar angle $\tilde{\theta}$ 
\begin{equation}
    \int_0^{\pi}\sin^2{\tilde{\theta}}d\tilde{\theta}e^{-iu\cos{\tilde{\theta}}}=\int_0^{\pi}d\tilde{\theta} e^{-iu\cos{\tilde{\theta}}}-\int_0^{\pi}\cos^2{\tilde{\theta}}d\tilde{\theta} e^{-iu\cos{\tilde{\theta}}} = {J_1(u) \over u} \ .
    \end{equation}
   
There remains to integrate over the radial variable $r$. Repeating the same steps of Section~\ref{Sec:RadialIntegral}, eventually one gets
\begin{equation}
     \widetilde{h}_{\mu\nu}(\vec{q})=-\frac{32\pi GM}{3}\int_{u_a}^\infty \frac{udu}{|\vec{q}|^2} K_\mu K_\nu \frac{J_1(u)}{u}\ ,
\end{equation}
where $u_a=|a_{ij}q^{j}| = |\vec{u}_a|$. Performing the integral over the variable $u$, along the lines described in Appendix~\ref{app:ExpCalc}, one finally gets
\begin{align}
\widetilde{h}_{00}(\vec{q}\,)=&-\frac{32\pi GM}{3|\vec{q}|^2}J_0(u_a)\ ,\\
     \widetilde{h}_{0i}(\vec{q}\,)=&i \frac{32 \pi GM}{3}\Bigg[\frac{q_i}{|\vec{q}|^3}\sqrt{\frac{\pi}{2u_a}}J_{1\over 2}(u_a)-\frac{a_{ik}q^{k}}{|\vec{q}|^2u_a}J_1(u_a)\Bigg]\ ,\\
    \widetilde{h}_{ij}(\vec{q}\,)=&-\frac{32\pi GM}{3}\Bigg[\frac{1}{|\vec{q}|^2}\left(\delta_{ij} -2\frac{q_iq_j}{|\vec{q}|^2}\right)\frac{J_1(u_a)}{u_a}-\frac{a_{ik}q^ka_{jh}q^h}{|\vec{q}|^2}\frac{J_2(u_a)}{u_a^2} \nonumber \\
&+\frac{1}{|\vec{q}|^3}\Big(a_{ik}q^kq_j+a_{jk}q^kq_i\Big)\sqrt{\frac{\pi}{2}}\frac{1}{(u_a)^{\frac{3}{2}}}J_{\frac{3}{2}}(u_a)\Bigg]\ .
\end{align}

Assembling the various pieces one gets the  `off-shell' scattering amplitude ($E'=E$, but $|\vec{p}'|\neq |\vec{p}|$ and $\vec{q}{\cdot}(\vec{p}+\vec{p}')\neq 0$)
\begin{equation}
\begin{aligned}
    i\mathcal{M}&=\frac{i32\pi GM}{3|\vec{q}\, |^2}\Bigg[E^2 J_0(u_a) 
-iE\Bigg(\frac{\vec{q}\cdot(\vec{p}\, '+\vec{p}\,)}{|\vec{q}\,|}j_{0}(u_a)-\frac{{\vec{u}_a}\cdot(\vec{p}+\vec{p}\, '\,)}{u_a}J_1(u_a)\Bigg) \\
&+\vec{p}\cdot\vec{p}\, '\frac{J_1(u_a)}{u_a}-\frac{2(\vec{q}\cdot\vec{p}\,)(\vec{q}\cdot\vec{p}\, '\,)}{|\vec{q}\, |^2}\frac{J_1(u_a)}{u_a}-({\vec{u}_a}\cdot\vec{p}\,)^2\frac{J_2(u_a)}{u_a^2} \\
&+\Big[({\vec{u}_a}\cdot\vec{p}\,)(\vec{q}\cdot\vec{p^\prime}\,)+({\vec{u}_a}\cdot\vec{p}\, '\,)(\vec{q}\cdot\vec{p}\,)\Big]\frac{1}{u_a}j_{1}(u_a)\Bigg],
\end{aligned}
\end{equation}
where we have used the relation between spherical Bessel and Bessel of the first kind ${j_\ell(z)=\sqrt{\frac{\pi}{2z}}J_{\ell+\frac{1}{2}}(z)}$.

On-shell, the expression of the scattering amplitude drastically simplifies. In particular all the terms with spherical Bessel functions drop and one gets
\begin{equation}
    i\mathcal{M}_{on-shell}=\frac{i32\pi GM}{3|\vec{q}\, |^2}\Bigg[E^2J_0(u_a)+iE\frac{{\vec{u}_a}\cdot{\vec{P}}}{u_a}J_1(u_a)+p^2\frac{J_1(u_a)}{u_a} -({\vec{u}_a}\cdot\vec{p}\,)^2\frac{J_2(u_a)}{u_a^2}\Bigg]\ .
\end{equation}

Thanks to Fermi golden rule, the elastic cross section can be computed from the rate divided by the  incoming energy flux ${Ev}/{V_4}$, where $V_4$ is the 4-dimensional (regulated) space `volume', 
\begin{equation}
    \frac{d\sigma}{d\Omega_3}=\frac{2\pi}{2Ev}|\mathcal{M}_{on-shell}|^2\delta(E-E^\prime)\frac{|\vec{p}\, '\,|^3d|\vec{p}\, '\,|}{(2\pi)^42E^\prime}\ .
\end{equation}

Using then $|\vec{p}\,|=Ev$ and integrating over $E^\prime$ yields
\begin{equation}
\frac{d\sigma}{d\Omega_3}=\frac{|\vec{p}\,|}{32\pi^3}|\mathcal{M}_{on-shell}|^2\ ,
\end{equation}
and the final very compact Rutherford-like formula reads 
\begin{equation}
\begin{aligned}
    \frac{d\sigma}{d\Omega_3}=&\frac{32G^2M^2}{3\pi^3}\frac{|\vec{p}\, |}{|\vec{q}\, |^4}\Bigg|E^2J_0(u_a)+iE\frac{{\vec{u}_a}\cdot{{\vec{P}}}}{u_a}J_1(u_a)+p^2\frac{J_1(u_a)}{u_a} \\
&    -\frac{2(\vec{q}\cdot\vec{p}\,)^2}{|\vec{q}|^2}\frac{J_1(u_a)}{u_a}-({\vec{u}_a}\cdot\vec{p}\,)^2\frac{J_2(u_a)}{u_a^2}\Bigg|^2 \ .
\end{aligned}
\end{equation}
Notice that the imaginary part of the on-shell amplitude is odd in the angular momentum $i.e.$ in the components of the vector $\vec{u}_a$.

\subsection{Eikonal expansion}

In order to study gauge-invariant gravitational observables such as the scattering angle and Shapiro time-delay, we can compute the eikonal phase, associated to the phase shift in the (elastic) scattering amplitude~\cite{DiVecchia:2023frv}. We start with a general discussion for MP BHs in arbitrary $D$ and then specialize to $D=5$ and eventually focus on the special cases $a_1=a_2$ and $a_2=0$.
Starting from the $S$-matrix, it is possible to write
\begin{equation}
S = 1 +iT = e^{2i\delta}
\end{equation}
that can be diagonalized in the partial wave basis as $S_\ell = e^{2i\delta_\ell (E)}$ or equivalently by a FT of the amplitude to impact parameter space. To lowest order 
\begin{equation}
T_1 = 2\delta_1 \ , \qquad  T_2 = 2\delta_2 +2{i}\delta_1^2 \ , \qquad ...
\end{equation}
In particular, at tree level one has to integrate $\vec{q}=\vec{p}-\vec{p}'$ transverse to $\vec{P}=\vec{p}+\vec{p}'$
\begin{equation}
2\delta(E, b)= \int {d^{D-2} q\over (2\pi)^{D-2}} e^{i\vec{q}{\cdot}\vec{b}} {\cal M}(E, \vec{p}, \vec{p}') = 
\int {d^{D-1} q\over (2\pi)^{D-2}} e^{i\vec{q}{\cdot}\vec{b}} \delta(\vec{q}\cdot {\vec{P}}) {\cal M}(E, \vec{p}, \vec{p}')\ .
\end{equation}
In some cases, it is convenient to replace $\delta(\vec{q}\cdot {\vec{P}})$ with an integral over an auxiliary variable $\xi$ and use 
\begin{equation} {\cal M}(E, \vec{p}, \vec{p}\, ') =\widetilde{h}_{\mu\nu}(\vec{q}\, ) p^\mu {p'}^\nu
\end{equation}
so that 
\begin{equation}
2\delta(E, b) = \int d\xi  {h}_{\mu\nu}(\vec{x}=\vec{b}+\xi{\vec{P}}) p^\mu {p'}^\nu\ .
\end{equation}

In other cases, one can observe that the Bessel functions appear in combinations of the form
\begin{equation}
{J_\alpha(u_a)\over u_a^\alpha} = \sum_n {(-1)^n \over n! \Gamma(n+\alpha+1)} {u_a^{2n}\over 2^{2n+\alpha}} \ ,
\end{equation}
where 
\begin{equation}
u_a^2 = q^i a_{ij}a^{jk}q_k \ ,
\end{equation}
so that one can compute the eikonal phase as the action of powers of the operator
\begin{equation}
u_a^2 = {\partial \over\partial b_i} a_{ij}a^{jk}{\partial \over\partial b^k} 
\end{equation} 
on the `basic' integral
\begin{equation}
\delta_0 = \int d^{D-2}q e^{iqb} {1\over q^2} = {C_D\over  b^{D-4}}\ ,
\end{equation}
and get a combination of terms of the form 
\begin{equation}
u_a^{2n} \delta_0 = \left({\partial \over\partial b_i} a_{ij}a^{jk}{\partial \over\partial b^k}\right)^{2n} {C_D\over  b^{D-4}}\ .
\end{equation}
In special cases it is possible to sum the series per each ${J_\alpha(u)/u^\alpha}$ term  in ${\cal M}_{on-shell}$). In particular for $a_i = a_j$ for all $i,j$ so that ${u_a^2= a^2 q^2}$, only the leading $u_a$ independent term contributes to long-distance interactions. All higher (even) powers of $u_a$ only give local terms in the amplitude.  Moreover, for consistency, one can check that for $a_i=0$ one reproduces the non-rotating case, {\it i.e.} Schwarzschild-Tangherlini BHs.

\subsection{$a_1=a_2$ case}

Let us consider the case in which $a_1=\pm a_2=a$. As already mentioned this case is particularly simple and interesting because the metric becomes cohomogeneity-one (with full rotational symmetry, despite the non vanishing angular momenta!) and all the higher gravitational multi-poles vanishes except for the current dipole and the stress quadrupole. 

In order to evaluate the eikonal phase one has to consider
\begin{equation}
    \delta=\frac{1}{2{P}}\int \frac{d^{\hat{d}}q}{(2\pi)^{\hat{d}}}e^{iq\cdot b}\mathcal{M}_{on-shell}\ ,
\end{equation}
with $\hat{d}=d-1=D-2$ and recall that $\vec b\cdot \vec{P}=\vec q\cdot \vec{P}=0$.  
Now for $a_1=a_2$ it turns out that $u_a=a q$, and relying on the master integral 
\begin{equation}
    \mathcal{F}(\hat{d}, \nu )=\int \frac{d^{\hat{d}}q}{(2\pi)^{\hat{d}}}e^{iq\cdot b}q^{2\nu}=\frac{2^{2\nu}}{\pi^{\hat{d}/2}}\frac{\Gamma(\nu +\hat{d}/2)}{\Gamma(-\nu )}\frac{1}{b^{2\nu +\hat{d}}}\ ,
\end{equation}
one can show that only the leading (zero$^{\rm th}$) order term in the series expansion of the Bessel functions contributes to the non-local part of the eikonal phase, namely the surviving contributions in the long-range limit $r\rightarrow+\infty$. Indeed, since 
\begin{equation}
    \frac{J_n(a q)}{(a q)^n}=\frac{1}{2^{n}n!}+O(a^2q^2)\ ,
\end{equation}
focusing only on the long-range physics one can write the eikonal phase as
\begin{equation}
    \delta=\frac{1}{2{P}}\int \frac{d^3q}{(2\pi)^3} e^{iq\cdot b}\frac{32 \pi G M}{3}\frac{1}{q^2}\Bigg(E^2+\tfrac{1}{8}{P}^2+\tfrac{i}{2}E\, \vec{u}_a{\cdot} \vec{P}-\tfrac{1}{32}(\vec{u}_a{\cdot}\vec{P})^2\Bigg)\ .
\end{equation}

Replacing now 
\begin{equation}
    \vec{u}_a{\cdot}\vec{P}\rightarrow-i {a}^{ij}{P}_{j}(\partial_b)_i\ ,
\end{equation}
applying such differential operator to the master integral $\mathcal{F}(\hat{d}=3, \nu=-1)=\tfrac{1}{4\pi b}$ one gets
\begin{equation}
    \vec u(\partial_b)\cdot \vec{P}\frac{1}{b}=\frac{i}{b^3}\vec{P}\cdot {a}\cdot \vec b\ ,\quad \Big(\vec u(\partial_b)\cdot \vec{P}\Big)^2\frac{1}{b}=-3\frac{(\vec{P}\cdot {a}\cdot \vec b)^2}{b^5}-\frac{\vec{P} \cdot {a}\cdot {a} \cdot \vec{P}}{b^3}\ ,
\end{equation}
and hence finally the eikonal phase reads
\begin{equation}
    \delta=\frac{4 G M }{3 {P} b}\Bigg(E^2+\tfrac{1}{8}{P}^2-\tfrac{1}{2}E\frac{{P}\cdot {a}\cdot b}{b^2}+\frac{1}{32}\Bigg(3\frac{({P}\cdot {a}\cdot b)^2}{b^4}+\frac{{P}\cdot {a}\cdot {a}\cdot {P}}{b^2}\Bigg)\Bigg)\ .
\end{equation}

We can clearly see how in the eikonal phase the leading dependence on the angular momentum $a$ of the MP BH is via the coupling with the angular momentum of the probe ${\ell}_{ij} \sim b_iP_j - b_j P_i$. The subleading terms are related to the stress quadrupole moment, since the mass quadrupole vanishes for this configuration. 

\subsection{$a_2=0$ case}

Let us now consider the case in which $a_1=a$ and $a_2=0$ which is important since it is the closest analogue in $D=5$ of the Kerr case in $D=4$. In order to simplify the calculations we are going to consider an equatorial scattering processes, namely we will take 
\begin{equation}
    \vec{P}=(P,0, 0, 0)\ ,\quad \vec b=(0, b, 0, 0)\ ,
\end{equation}

Considering the eikonal integral introduced in~\cite{Bianchi:2023lrg}
\begin{equation}
    \delta=-\frac{1}{2}\int_{-\infty}^{+\infty}d\xi h_{\mu\nu}(\vec x=\vec b+\xi \vec{P})p^\mu p^\nu\ ,
\end{equation}
performing the integral for the above kinematical situation one gets
\begin{equation}
    \delta=\frac{GM }{3 a^2}\Bigg(-b{P} +4a E+\frac{(b{P}-2aE)^2}{{P}\sqrt{b^2-a^2}}\Bigg)\ ,
\end{equation}
for $b>a$. Despite appearance, the leading term for $b^2\gg GM, a$ goes as 
\begin{equation}
 \delta = \frac{4 G M }{3 P b}\Bigg(E^2+\tfrac{1}{8}P^2-\frac{a P E}{2b}+\frac{a^2(3P^2+16E^2)}{32b^2}+{\cal O}\Big(\frac{1}{b^3}\Big)\Bigg)  \ , 
\end{equation}
which up to factors coincides with the leading result in the case $a_1=\pm a_2$, both reproduce the leading eikonal for non-rotating  Schwarzschild(-Tangherlini) BH, being independent of the angular momentum of the MP BH.

Another interesting kinematical configuration for $a_2=0$ is the case of a probe incident with $\vec{P}$ along the $(x_2,y_2)$ plane orthogonal to the plane $(x_1,y_1)$ with non-trivial angular momentum. Without loss of generality one can take $\vec{P}=(0,0,0,P)$ and integrate over $\vec{q}=(q_1,q_2,q_\perp,0)$ orthogonal to $\vec{P}$ and with $q_1^2+q_2^2=q_\parallel^2$, where parallel and orthogonal are intended with respect to the rotational place $(x_1, y_1)$. Switching to polar coordinates $(q_\parallel,\varphi_\parallel)$ in the $(q_1,q_2)$ plane and taking into account that $\vec{u}_a=(a q_2,-a q_1,0,0)$ so that ${u_a^2 =a^2q_\parallel^2}$,  one has 
\begin{equation}
\delta =  \frac{2 \pi G M}{3P\pi^3} \int \frac{q_\parallel dq_\parallel d\varphi_\parallel dq_\perp}{q_\parallel^2+q_\perp^2}e^{iq_\parallel b_\parallel\cos\varphi_\parallel + i q_\perp b_\perp} \Bigg(E^2 J_0(u_a)+\tfrac{1}{4}{P}^2\frac{J_1(u_a)}{u_a}\Bigg)  \ ,
\end{equation}
since in this configuration $\vec{u}_a\cdot \vec P=0$.

The angular integral produces a Bessel function 
\begin{equation}
\int_0^{2\pi} d\varphi_\parallel e^{iq_\parallel b_\parallel\cos\varphi_\parallel} = 2\pi J_0(q_\parallel b_\parallel) \ ,
\end{equation}
while the integral over $q_\perp$ can be computed with the method of residues 
\begin{equation}
\int_{-\infty}^{+\infty} {dq_\perp \over q_\parallel^2+q_\perp^2}e^{i q_\perp b_\perp} =
{\pi \over q_\parallel} e^{-q_\parallel|b_\perp|} \ ,
\end{equation}
leading to the expression 
\begin{equation}
    \delta=\frac{4GM}{3P}\int_0^{+\infty}dq_\parallel J_0(b_\parallel q_\parallel)e^{-q_\parallel|b_\perp|}\Bigg(E^2 J_0(a q_\parallel)+\tfrac{1}{4}P^2\frac{J_1(a q_\parallel)}{a q_\parallel}\Bigg)\ .
\end{equation}
One is then left with integrals of the form
\begin{equation}
\int_0^\infty  dq e^{-q_\parallel|b_\perp|} J_0(q_\parallel b_\parallel) {J_n(q_\parallel a) \over (q_\parallel a)^n}
\end{equation}
with $n=0,1$ that can be mapped into the known integral~\cite{Gradshteyn:1943cpj}
\begin{equation}
\label{eq:masterintegral626}
\begin{aligned}
    \mathcal{I}(\lambda; \mu,\nu; \alpha, \beta, \gamma)=&\int_0^{+\infty} dx\,  x^{\lambda-1}e^{-\alpha x} J_\mu(\beta x)J_\nu(\gamma x)=\frac{\beta^\mu\gamma^\nu}{\Gamma(\nu+1)}2^{-\nu-\mu}\alpha^{-\lambda-\mu-\nu}\\
    &\times\sum_{m=0}^{+\infty}\frac{\Gamma(\lambda+\mu+\nu+2m)}{m!\Gamma(\mu+m+1)}{}_2F_1(-m, -\mu-m;\nu+1;\tfrac{\gamma^2}{\beta^2})\Big(-\frac{\beta^2}{4\alpha^2}\Big)^m\ ,
\end{aligned}
\end{equation}
so that the eikonal phase reads
\begin{equation}\label{eq:eikonalorthogonal}
    \delta=\frac{4GM}{3P}\Bigg(E^2 \mathcal{I}(1, 0, 0, |b_\perp|, b_\parallel, a)+\tfrac{P^2}{4a}\mathcal{I}(0, 0, 1, |b_\perp|, b_\parallel, a)\Bigg)\ .
\end{equation}

Notice that ${}_2F_1(-m, -\mu-m;\nu+1;z)$ above are (Jacobi) polynomials of degree $m$. Moreover in the special case in which $\lambda=1$ and $\mu=\nu=\sigma$ the integral in \eqref{eq:masterintegral626} simplifies into
\begin{equation}
\mathcal{I}(1; \sigma,\sigma; |b_\perp|, b_\parallel, a)=\int_0^\infty dq_\parallel e^{-|b_\perp| q_\parallel} J_\sigma(q_\parallel b_\parallel) J_\sigma(q_\parallel a) = {1\over \pi \sqrt{ab_\parallel} } Q_{\sigma-{1\over2}} \left( {a^2 + b^2 \over 2 ab_\parallel}\right)
\end{equation}
where $b^2=b_\parallel^2+b_\perp^2$ and  ${Q_{\mu}(x)}$ are the associated Legendre functions of the second kind, that
can be expressed in terms of hypergeometric functions as
\begin{equation}
 Q^{\mu}_{\nu} (x) = e^{i\pi \mu} {\pi^{1/2} \Gamma(\mu+\nu+1) (x^2-1)^{\mu/2} \over x^{\mu+\nu+1} \Gamma\left(\nu + {3\over 2}\right)}    {}_2F_1\left({\mu+\nu \over 2} +1, {\mu+\nu +1\over 2} ; \nu + {3\over 2}; {1\over x^2}\right)             
\end{equation}
In particular for $\mu=0$ one has $Q_\nu(x) = Q^0_\nu(x)$, so that for $\sigma= 0$ (\textit{i.e.} $\nu=-1/2$) one finds
\begin{equation}
{1\over \pi \sqrt{ab_\parallel}} Q_{-{1\over2}} \left( {a^2 + b^2 \over 2 ab_\parallel}\right) = 
{\sqrt{2} \over \sqrt{a^2 + b^2  }} {}_2F_1\left({3\over 4}, {1\over 4} ; 1; { 4 a^2b_\parallel^2 \over (a^2 + b^2 )^2 }\right) \ .
\end{equation} 
Expanding Eq. \eqref{eq:eikonalorthogonal} for $a\ll b_\parallel,|b_\perp|$ one has 
\begin{equation}
 \delta=\frac{4GM}{3 Pb } \left(E^2 + {1\over 8} P^2\right)\left(1  + a^2{(b^2-3b_\perp^2) \over 8b^4}\frac{P^2+16E^2}{P^2+8E^2}  + {\cal O}(a^4)\right)      
\end{equation}
where the leading term reproduces the spherically symmetric case.
It is interesting to compare this result with the four-dimensional scattering off a Kerr BH with incident momentum parallel to the axis of rotation in which one finds
\begin{equation}
\delta_{\rm Kerr} = -{G M \over P} \left(E^2+{P^2 \over 4}\right)\log { \mu^2|\vec{b}|^2} \ ,
\end{equation}
with $\mu$ an IR regulator and no dependence on $a$ for this very special kinematic configuration, so that the eikonal phase becomes effectively equal to the non-rotating case.
The important result then is the fact that in dimensions higher than four, in the limit of axial incidence of the probe there is still  an angular momentum dependence left in the expression of the eikonal phase, differently with respect to the Kerr case. Therefore,  even though in this configuration the stress multipoles vanish, this difference in phenomenology proves once again how going to higher dimensions enriches gravitational physics due to geometrical arguments.

\section{Conclusions and outlook}\label{sec:Conclusion}

Let us conclude by summarizing our results and comment on possible directions for future investigation. Relying on our previous analysis for the scattering of massive (neutral) scalar probes off Kerr(-Newman) BHs in the {KS} gauge ~\cite{Bianchi:2023lrg}, we have tackled the same problem for Myers-Perry (MP) BHs in arbitrary  $D=d+1$ dimensions~\cite{Myers:1986un}. 
It is worth recalling that Myers-Perry BHs are the unique stationary, non-extremal,
asymptotically flat, vacuum BH solutions with spherical topology of the horizon~\cite{Morisawa:2004tc}.

Working in the KS gauge, whereby there is only a tri-linear coupling with the graviton, we arrived at very compact and explicit formulae for the tree-level (elastic) scattering amplitude and the cross-section for arbitrary angular momentum in the probe approximation $m\ll M$.  We highlighted the main differences between $D=d+1$ even or odd in the intermediate steps, even though the final formulae have the same structure with the effect of the BH spin coded in Bessel functions $J_\alpha(u_a)$, $J_{\alpha{+}1}(u_a)$ and $J_{\alpha{-}1}(u_a)$ where $\alpha=(D{-}3)/2$ and $u_a=|\vec{u}_a|=|a\vec{q}|$ with $a_{ij}$ the spin matrix of the MP BH and $\vec{q}=  \vec{p}- \vec{p}'$ the transferred momentum. The Bessel functions are exactly the same as those appearing in the form factors that encode the multipolar structure of MP BHs~\cite{Bianchi:2024shc}.
In order to render more intuitive our results and highlight the role of the `stress' multipoles~\cite{Gambino:2024uge}, we have then focused on the $D=4+1$ case and explicitly computed the leading eikonal phase for special choices of the angular momenta, {\it viz.} $a_1=\pm a_2=a$ and $a_2=0$, and of the kinematics. 
In principle one can consider higher-order `loop' corrections to the amplitude in the KS gauge and then to the eikonal phase. Very much as in the Kerr case in $D=4$~\cite{Bianchi:2023lrg}, we expect the Bessel functions with `wrong' index, \textit{i.e.}  $\alpha\pm 1/2$, that don't contribute on-shell, to produce the necessary off-shell terms. This allows to organize the amplitude in an all-loop closed formula, thanks to the fact that in the KS gauge the $\ell$-loop amplitude is captured by a single comb-like (ladder) diagram. We expect cancellation of the denominator in the propagator and plan to investigate the case $D=5$ in the future since the analysis should be easier thanks to the absence of square roots.
Another interesting direction of investigation, already mentioned in the introduction, is to compare `our' results with those for the scattering of massive neutral (scalar) probes off ECO's or fuzzballs or other BH mimickers~\cite{Cardoso:2019rvt, Gambino:2025xdr}. 

The fuzzball proposal~\cite{Lunin:2001jy, Bena:2007kg, Skenderis:2008qn, Bianchi:2017sds, Bianchi:2022qph, DiRusso:2024hmd}, motivated by string theory, posits that a large fraction of the micro-states be smooth and horizonless geometries. However due to a no-go theorem by (lo and behold) Einstein and Pauli~\cite{Gibbons:2013tqa, Einstein:1941xxx, Einstein:1943ixi}, the proposal cannot work in 4-d, unless one includes other fields (vectors, scalars, ...). This suggests to look for a higher dimensional origin of the micro-structure at the scale of the would-be horizon. The lowest possible dimension is $D=5$ where almost 40 years ago Strominger and Vafa~\cite{Strominger:1996sh} gave a string theoretic interpretation of the micro-state counting of a special class of charged BPS BHs in $D=5$.
Later on explicit micro-states have been constructed and studied in some detail, also from the 4-dimensional perspective~\cite{Bena:2015bea, Bena:2016ypk}. More recently an extremely simple class of smooth horizonless solutions of 5-d Einstein-Maxwell equations has been constructed by Bah and Heidmann and named Top(ological) Stars~\cite{Bah:2020pdz}. They can be considered as `caricature' of fuzzballs~\cite{Bianchi:2023sfs, Heidmann:2023ojf, Bianchi:2024vmi, Cipriani:2024ygw, Bena:2024hoh, Dima:2024cok, DiRusso:2025lip, Bianchi:2024rod} . A rotating version has been proposed very recently and dubbed Rotating Top(ological) Stars~\cite{Bianchi:2025uis}. In order to probe these and other interesting higher-dimensional compact objects, scattering processes \`a la Rutherford are once again the preferred choice. 

\begin{acknowledgments}
We  thank  D.~Bini, A. Cipriani, G. Dibitetto, G.~Di~Russo, R.~Emparan, M.~Firrotta, F.~Fucito, R.~Gonzo,  J.~F.~Morales, P.~Pani and A.~Ruiperez Vicente for useful discussions and comments. 
C.~G. acknowledges the hospitality of the IFAE (Institut de Física d'Altes Energies) and Universitat de Barcelona/ICCUB during the time this project was finalized.
This work is partially supported by Sapienza University of Rome (``Progetti per Avvio alla Ricerca - Tipo 1'',
protocol number AR1241906DC8FF32), the MIUR PRIN contract 2020KR4KN2 ``String Theory as a bridge
between Gauge Theories and Quantum Gravity'' and the INFN project ST\&FI ``String Theory \& Fundamental Interactions''.
\end{acknowledgments}

\appendix

\section{Explicit calculation of $\widetilde{h}_{\mu\nu}$ in $D$ dimensions}
\label{app:ExpCalc}
In this appendix we give the details for the computation of the radial integrals that appear in the FT of ${h}_{\mu\nu}$. 
We give them in the general case of the $D$-dimensional MP BHs.
Let us start from $\widetilde{h}_{00}$ reading 
\begin{equation}
\begin{aligned}
    \widetilde{h}_{00}&=C_D\int_{u_a}^\infty\frac{udu}{|\vec{q}\,|^2}\frac{J_{\alpha}(u)}{u^{\alpha}}=\frac{C_D}{|\vec{q}\,|^2}\frac{J_{{\alpha}-1}(u_a)}{u_a^{{\alpha}-1}}\ ,
\end{aligned}
\end{equation}
where $\alpha = (D-3)/2$ and the recursion relations for Bessel integrals has been used
\begin{equation}
    \int_{z_0}^\infty z^{1-{\alpha}}J_{\alpha}(z) dz =z_0^{-{\alpha}+1}J_{{\alpha}-1}(z_0) .
\end{equation}
Passing to $\widetilde{h}_{0i}$, one can consider $\widetilde{h}_{01}$ for definiteness and then generalize to the other components by symmetry arguments. In order to compute the radial integral, one has to express $K_1$ using the generalization to $D$ dimensions of \eqref{eq:K nella rappr degli impulsi} that yields
\begin{equation}
\begin{aligned}
    \widetilde{h}_{01}&=C_D\int_{u_a}^\infty\frac{iudu}{|\vec{q}\,|^2}\Bigg(\frac{r\partial_{q_{x_1}}+a_1\partial_{q_{y_1}}}{r^2+a_1^2}\Bigg)\frac{J_{\alpha}(u)}{u^{\alpha}} \\
    &=C_D\Bigg[\int_{u_a}^\infty \frac{iudu}{|\vec{q}\,|^2}\frac{rq_{x_1}}{u}\frac{d}{du}\Bigg(\frac{J_{\alpha}(u)}{u^{\alpha}}\Bigg)+\int_{u_a}^\infty \frac{iudu}{|\vec{q}|^2}\frac{a_1q_{y_1}}{u}\frac{d}{du}\Bigg(\frac{J_{\alpha}(u)}{u^{\alpha}}\Bigg)\Bigg]\ ,
\end{aligned}
\end{equation}
where \eqref{eq:azione di K su una fun di u} has been used.
Keeping in mind that $r=\sqrt{u^2-u_a^2}/|\vec{q}\,|$, the first integral can be computed by parts 
\begin{equation}
    \frac{iC}{|\vec{q}\,|^2}rq_{x_1}\frac{J_{\alpha}(u)}{u^{\alpha}}\Bigg|_{u_a}^\infty-C_D\int_{u_a}^\infty \frac{idu}{|\vec{q}\,|^3}\frac{uq_{x_1}}{\sqrt{u^2-u_a^2}}\frac{J_{\alpha}(u)}{u^{\alpha}} =-C_D\int_0^\infty\frac{ids}{|\vec{q}\,|^3}q_{x_1}\frac{J_{\alpha}(\sqrt{s^2+u_a^2})}{(s^2+u_a^2)^{\frac{{\alpha}}{2}}}\ ,
\end{equation}
The boundary term gives no contribution and we have changed variable to $s^2=u^2-u_a^2$.
We now need two formulae to proceed. First, the integral 
\begin{equation}
\label{eq:int notevole per le bessel mod}
    \int_0^{\infty}du(u^2+\beta^2)^{-\frac{{\alpha}}{2}}J_{\alpha}(\gamma\sqrt{u^2+\beta^2}) \\
    =\sqrt{\frac{\pi}{2}}\gamma^{-{\alpha}}\beta^{\frac{1}{2}-{\alpha}}(\gamma^2)^{\frac{{\alpha}}{2}-\frac{1}{4}}J_{{\alpha}-\frac{1}{2}}(\beta\gamma)\ ,
\end{equation}
which is tabulated in~\cite{Gradshteyn:1943cpj}\footnote{A similar integral with modified Bessel functions $K_\alpha$ is  given in the appendix of~\cite{berg:2022feq} {\it viz.} $$\int_0^{\infty}du(u^2+\beta^2)^{-\frac{{\alpha}}{2}}K_{\alpha}(\gamma\sqrt{u^2+\beta^2}) \\
    =\sqrt{\frac{\pi}{2}}\gamma^{-{\alpha}}\beta^{\frac{1}{2}-{\alpha}}(\gamma^2)^{\frac{{\alpha}}{2}-\frac{1}{4}}K_{{\alpha}-\frac{1}{2}}(\beta\gamma)\ .$$}.
The second integral is immediate and summed to the first yields
\begin{equation}
    \widetilde{h}_{01}=-C_D\Bigg[\frac{iq_{x_1}}{|\vec{q}\,|^3}\sqrt{\frac{\pi}{2}}\frac{J_{{\alpha}-\frac{1}{2}}(u_a)}{u_a^{{\alpha}-\frac{1}{2}}}+\frac{1}{|\vec{q}\,|^2}a_1q_{y_1}\frac{J_{\alpha}(u_a)}{u_a^{\alpha}}\Bigg]\ ,
\end{equation}
that can be generalized by symmetry to
\begin{equation}
    \widetilde{h}_{0i}=-C_D\Bigg[\frac{iq_{i}}{|\vec{q}\,|^3}\sqrt{\frac{\pi}{2}}\frac{J_{{\alpha}-\frac{1}{2}}(u_a)}{u_a^{{\alpha}-\frac{1}{2}}}+\frac{1}{|\vec{q}\,|^2}a_{ij}q^{j}\frac{J_{\alpha}(u_a)}{u_a^{\alpha}}\Bigg]\ .
\end{equation}

The component $K_{z}=\frac{z}{r}$ looks different from the rest but the peculiarity of these terms (present only for $D$ even) is taken into account in the term $a_{ij}q^j$, that vanishes for $i$  corresponding to $z$, leaving only the relevant term.
The remaining $\widetilde{h}_{ij}$ are similar. One starts with  $\widetilde{h}_{11}$ and then generalizes to the other (diagonal and mixed) components by symmetry. Integrating by parts the relevant integral can be decomposed as follows
\begin{equation}
\begin{aligned}
    \widetilde{h}_{11}&=-C_D\int_{u_a}^\infty\frac{udu}{|\vec{q}\,|^2}\Bigg(\frac{r\partial_{q_{x_1}}+a_1\partial_{q_{y_1}}}{r^2+a_1^2}\Bigg)\Bigg(\frac{r\partial_{q_{x_1}}+a_1\partial_{q_{y_1}}}{r^2+a_1^2}\Bigg)\frac{J_{\alpha}(u)}{u^{\alpha}} \\
   & =-C_D\int_{u_a}^\infty\frac{udu}{|\vec{q}\,|^2}\Bigg[\frac{r^2+a_1^2}{r^2+a_1^2}\frac{1}{u}\frac{d}{du}\Bigg(\frac{J_{\alpha}(u)}{u^{\alpha}}\Bigg)-(rq_{x_1}+a_1q_{y_1})^2\frac{1}{u^3}\frac{d}{du}\Bigg(\frac{J_{\alpha}(u)}{u^{\alpha}}\Bigg) \\
    &+(rq_{x_1}+a_1q_{y_1})^2\frac{1}{u^2}\frac{d^2}{du^2}\Bigg(\frac{J_{\alpha}(u)}{u^{\alpha}}\Bigg)\Bigg]\ .
\end{aligned}
\end{equation}
The first integral is immediate, the other two contain  three terms each with different powers of $r$ in the integrand:  $r^2$, $r$ and 1.
Start with the ones with $r^2 = (u^2-u_a^2)/|\vec{q}\,|^2$ and 1 we get
\begin{equation}
    -\int_{u_a}^\infty\frac{du}{|\vec{q}\,|^4}q_{x_1}^2\frac{d}{du}\Bigg(\frac{J_{\alpha}(u)}{u^{\alpha}}\Bigg)+\int_{u_a}^\infty\frac{du}{|\vec{q}\,|^4}\frac{u_a^2q_{x_1}^2-a_1^2q_{y_1}^2|\vec{q}\,|^2}{u^2}\frac{d}{du}\Bigg(\frac{J_{\alpha}(u)}{u^{\alpha}}\Bigg)\ ,
\end{equation}
and  
\begin{equation}
\begin{aligned}
    &\int_{u_a}^\infty\frac{du}{|\vec{q}\,|^4}uq_{x_1}^2\frac{d^2}{du^2}\Bigg(\frac{J_{\alpha}(u)}{u^{\alpha}}\Bigg)-\int_{u_a}^\infty\frac{du}{|\vec{q}\,|^4}\frac{u_a^2q_{x_1}^2-a_1^2q_{y_1}^2|\vec{q}\,|^2}{u}\frac{d^2}{du^2}\Bigg(\frac{J_{\alpha}(u)}{u^{\alpha}}\Bigg) \\
    &=\frac{uq_{x_1}^2}{|\vec{q}\,|^4}\frac{d}{du}\Bigg(\frac{J_{\alpha}(u)}{u^{\alpha}}\Bigg)\Bigg|_{u_a}^\infty-\int_{u_a}^\infty \frac{du}{|\vec{q}\,|^4}q_{x_1}^2\frac{d}{du}\Bigg(\frac{J_{\alpha}}{u^{\alpha}}\Bigg) \\
    &-\frac{u_a^2q_{x_1}^2-a_1^2q_{y_1}^2|\vec{q}\,|^2}{|\vec{q}\,|^4}\frac{1}{u}\frac{d}{du}\Bigg(\frac{J_{\alpha}(u)}{u^{\alpha}}\Bigg)\Bigg|_{u_a}^\infty-\int_{u_a}^\infty\frac{du}{|\vec{q}\,|^4}\frac{u_a^2q_{x_1}^2-a_1^2q_{y_1}^2|\vec{q}\,|^2}{u^2}\frac{d}{du}\Bigg(\frac{J_{{\alpha}}(u)}{u^{\alpha}}\Bigg)\ .
\end{aligned}
\end{equation}
Integrating by parts, using the derivative relation for Bessel functions 
\begin{equation}
    \Bigg(\frac{1}{u}\frac{d}{du}\Bigg)^k(u^{-{\alpha}}J_{{\alpha}}(u))=(-1)^ku^{-{\alpha}-k}J_{{\alpha}+k}(u)\ ,
\end{equation}
and summing all the terms yields
\begin{equation}
    \frac{2}{|\vec{q}\,|^4}q_{x_1}^2\frac{J_{\alpha}(u_a)}{u_a^{\alpha}}+\frac{a_1^2q_{y_1}^2|\vec{q}\,|^2}{|\vec{q}\,|^4}\frac{J_{{\alpha}+1}(u_a)}{u_a^{{\alpha}+1}}\ .
\end{equation}
One is left with the integrals involving terms in $r$  
\begin{equation}
    -\int_{u_a}^\infty\frac{du}{|\vec{q}\,|^2}2ra_1q_{x_1}q_{y_1}\frac{1}{u^2}\frac{d}{du}\Bigg(\frac{J_{\alpha}(u)}{u^{\alpha}}\Bigg)\ ,
\end{equation}
and
\begin{equation}
\begin{aligned}
   &\int_{u_a}^\infty \frac{du}{|\vec{q}\,|^2}2ra_1q_{x_1}q_{y_1}\frac{1}{u}\frac{d^2}{du^2}\left(\frac{J_\alpha(u)}{u^\alpha}\right)= \frac{2ra_1q_{x_1}q_{y_1}}{|\vec{q}\,|^2}\frac{1}{u}\frac{d}{du}\Bigg(\frac{J_{\alpha}(u)}{u^{\alpha}}\Bigg)\Bigg|_{u_a}^\infty \\
    &+\int_{u_a}^\infty \frac{du}{|\vec{q}\,|^2}2ra_1q_{x_1}q_{y_1}\frac{1}{u^2}\frac{d}{du}\Bigg(\frac{J_{\alpha}(u)}{u^{\alpha}}\Bigg)-\int_{u_a}^\infty\frac{du}{|\vec{q}\,|^3}\frac{2a_1q_{x_1}q_{y_1}}{\sqrt{u^2-u_a^2}}\frac{d}{du}\Bigg(\frac{J_{\alpha}(u)}{u^{\alpha}}\Bigg)\ .
\end{aligned}
\end{equation}

The boundary terms vanish and summing the two integrals one is left with an integral of the form \eqref{eq:int notevole per le bessel mod} that yields
\begin{equation}
    \frac{2a_1q_{x_1}q_{y_1}}{|\vec{q}|^3}\sqrt{\frac{\pi}{2}}\frac{J_{{\alpha}+\frac{1}{2}}(u_a)}{u_a^{{\alpha}+\frac{1}{2}}}\ .
\end{equation}
The final result reads
\begin{equation}
    \widetilde{h}_{11}=C_D\Bigg[\frac{1}{|\vec{q}\,|^2}\frac{J_{\alpha}(u_a)}{u_a^{\alpha}}-\frac{2q_{x_1}^2}{|\vec{q}\,|^4}\frac{J_{\alpha}(u_a)}{u_a^{\alpha}}-\frac{a_1^2q_{y_1}^2}{|\vec{q}\,|^2}\frac{J_{{\alpha}+1}(u_a)}{u_a^{{\alpha}+1}}-\frac{2a_1q_{x_1}q_{y_1}}{|\vec{q}\,|^3}\sqrt{\frac{\pi}{2}}\frac{J_{{\alpha}+\frac{1}{2}}(u_a)}{u_a^{{\alpha}+\frac{1}{2}}}\Bigg]\ ,
\end{equation}
that can be generalized to the other components (by symmetry and tracelessness of $\widetilde{h}_{\mu\nu}$) as 
\begin{equation}
    \widetilde{h}_{ij}=C_D\Bigg[\frac{\delta_{ij}}{|\vec{q}\,|^2}\frac{J_{\alpha}(u_a)}{u_a^{\alpha}}-\frac{2q_{i}q_{j}}{|\vec{q}\,|^4}\frac{J_{\alpha}(u_a)}{u_a^{\alpha}}-\frac{a_{ik}q^k a_{jh}q^h}{|\vec{q}\,|^2}\frac{J_{{\alpha}+1}(u_a)}{u_a^{{\alpha}+1}}-\frac{a_{ik}q^k q_j + a_{jk}q^k q_i}{|\vec{q}\,|^3}\sqrt{\frac{\pi}{2}}\frac{J_{{\alpha}+\frac{1}{2}}(u_a)}{u_a^{{\alpha}+\frac{1}{2}}}\Bigg]\ .
\end{equation}

\section{Jacobian for the transformation from cartesian to poly-spherical coordinates in D dimensions} \label{app:Jacobian}
Let us compute the Jacobian for the change of coordinates used in Section~\ref{sec:MP}. Clearly we deal with the Jacobian for the spatial coordinates {\it i.e.}  $d= D-1$ variables. We will follow the approach in~\cite{Muleshkov:2016easy}, applying it to our poly-spherical coordinates. We need to distinguish between $D$ odd from $D$ even since the Jacobian is slightly different.

\subsection{D odd, d even case}
In the odd $D$ case, the Jacobian matrix reads
\begin{equation}
{\cal J}_{ij}=\begin{pmatrix}
 \frac{\partial x_1}{\partial r} & \frac{\partial x_1}{\partial \mu_1} & \frac{\partial x_1}{\partial \phi_1} & 0 & 0 & ... & ... & 0  \\
 \frac{\partial y_1}{\partial r} & \frac{\partial y_1}{\partial \mu_1} & \frac{\partial y_1}{\partial \phi_1} & 0 & 0 & ... & ... & 0  \\
 \frac{\partial x_2}{\partial r} & 0 & 0 & \frac{\partial x_2}{\partial \mu_2} & \frac{\partial x_2}{\partial \phi_2} & ... & ... & 0  \\
  \frac{\partial y_2}{\partial r} & 0 & 0 & \frac{\partial y_2}{\partial \mu_2} & \frac{\partial y_2}{\partial \phi_2} & ... & ... & 0  \\
  \vdots & \vdots & \vdots & \vdots & \vdots & \vdots  & \vdots & \vdots \\
  \frac{\partial x_n}{\partial r} & \frac{\partial x_n}{\partial \mu_1} & 0 & \frac{\partial x_n}{\partial \mu_2} & 0 & ... & ... & \frac{\partial x_n}{\partial \phi_n} \\
  \frac{\partial y_n}{\partial r} & \frac{\partial y_n}{\partial \mu_1} & 0 & \frac{\partial y_n}{\partial \mu_2} & 0 & ... & ...  & \frac{\partial y_n}{\partial \phi_n} \\
\end{pmatrix}\ .
\end{equation}
The variables are $r$, all the $\phi_k$ and the $\mu_k$ except for $\mu_n$ that is constrained by \eqref{eq:cond sui mui}.
We notice that apart from the first column and the last two rows we have a block diagonal matrix. We then collect the first term in each row and expand the determinant using Laplace method with respect to the first column. For the terms of the expansion, except for the two obtained from the last rows, one has the possibility of expanding the row related to the expansion index and, then, the first column of the remaining matrix. Proceeding in this way we obtain two terms which multiply the determinant of a block matrix in which the blocks relating to the indices $n$ and the expansion index (related to the first Laplace expansion that has been done) are missing. For the last two terms, however, it is enough to expand with respect to the last column and then get the determinant of the block diagonal matrix with all the terms. We finally get 
\begin{equation}
\begin{aligned}
{\cal J}_{o}&=\Bigg(\prod_{k=1}^{\frac{D-1}{2}}\frac{\partial x_k }{\partial r }\frac{\partial y_k}{\partial r}\Bigg)\Bigg[\sum_{l=1}^{\frac{D-3}{2}}\Big(-\frac{\partial y_l}{\partial \phi_l}\Big(\frac{\partial y_l }{\partial r}\Big)^{-1}+\frac{\partial x_l}{\partial \phi_l}\Big(\frac{\partial x_l}{\partial r}\Big)^{-1}\Big) \\
&\times\Bigg[ (-1)^{D-1}\frac{\partial x_n}{\partial \mu_l}\Big(\frac{\partial x_n}{\partial r}\Big)^{-1}\frac{\partial y_n}{\partial \phi_n}\Big(\frac{\partial y_n}{\partial r}\Big)^{-1}+(-1)^{D-2}\frac{\partial x_n}{\partial \phi_n}\Big(\frac{\partial x_n}{\partial r}\Big)^{-1}\frac{\partial y_n}{\partial \mu_l}\Big(\frac{\partial y_n}{\partial r}\Big)^{-1}\Bigg] \\
&\times \prod_{k\neq i}^{\frac{D-3}{2}}\Bigg(\frac{\partial x_k}{\partial \mu_k}\Big(\frac{\partial x_k}{\partial r}\Big)\frac{\partial y_k}{\partial \phi_k}\Big(\frac{\partial y_k}{\partial r}\Big)^{-1}-\frac{\partial x_k}{\partial \phi_k}\Big(\frac{\partial x_k}{\partial r}\Big)^{-1}\frac{\partial y_k}{\partial \mu_k}\Big(\frac{\partial y_k}{\partial r}\Big)^{-1}\Bigg) \\
&+\Bigg[(-1)^{D-1}\frac{\partial y_N }{\partial \phi_n }\Big(\frac{\partial y_n }{\partial r }\Big)^{-1}+(-1)^{D-2}\frac{\partial x_n }{\partial \phi_n }\Big(\frac{\partial x_n }{\partial r }\Big)^{-1}\Bigg] \\
&\times \prod_{k=1}^{\frac{D-3}{2}}\Bigg(\frac{\partial x_k}{\partial \mu_k}\Big(\frac{\partial x_k}{\partial r}\Big)^{-1}\frac{\partial y_k}{\partial \phi_k}\Big(\frac{\partial y_k}{\partial r}\Big)^{-1}-\frac{\partial x_k}{\partial \phi_k}\Big(\frac{\partial x_k}{\partial r}\Big)^{-1}\frac{\partial y_k}{\partial \mu_k}\Big(\frac{\partial y_k}{\partial r}\Big)\Bigg)^{-1}\Bigg]\ .
\end{aligned}
\end{equation}

The explicit expressions of the partial derivatives read
\begin{align}
    \frac{\partial x_k}{\partial r}&=\frac{r}{\sqrt{r^2+a_k^2}}\mu_k\cos{\phi_k}\ , & \frac{\partial y_k}{\partial r}&=\frac{r}{\sqrt{r^2+a_k^2}}\mu_k\sin{\phi_k}\ ,\\
    \frac{\partial x_k}{\partial \mu_k}&=\sqrt{r^2+a_k^2}\cos{\phi_k}\ , & \frac{\partial y_k}{\partial \mu_k}&=\sqrt{r^2+a_k^2}\sin{\phi_k}\ ,\\
    \frac{\partial x_k}{\partial \phi_k}&=-\sqrt{r^2+a_k^2}\mu_k\sin{\phi_k}\ , & \frac{\partial y_k}{\partial \phi_k}&=\sqrt{r^2+a_k^2}\mu_k\cos{\phi_k}\ ,\\
    \frac{\partial x_n}{\partial \mu_k}&=-\sqrt{r^2+a_n^2}\frac{\mu_k}{\mu_n}\cos{\phi_n}\ , & \frac{\partial y_n}{\partial \mu_k}&=-\sqrt{r^2+a_n^2}\frac{\mu_k}{\mu_n}\sin{\phi_n}\ .
\end{align}
Replacing these in the expression for ${\cal J}$ above, collecting the terms and multiplying by the external factor we get
\begin{equation}
    {\cal J}_{o}=(-1)^{D-1}r\Bigg(\prod_{k\neq n}^{\frac{D-1}{2}}\mu_k\Bigg)\Bigg(\prod_{k=1}^{\frac{D-1}{2}}(r^2+a_k^2)\Bigg)\Bigg(\sum_{k=1}^{\frac{D-1}{2}}\frac{\mu_k^2}{r^2+a_k^2}\Bigg)\ ,
\end{equation}
or more compactly
\begin{equation}
    {\cal J}_{o}=\Bigg(\prod_{k\neq n}^{\frac{D-1}{2}}\mu_k\Bigg){\Pi F \over r}\ .
\end{equation}

where $F$ has been rewritten as 
\begin{equation}
    F=1-\sum_{k=1}^{\frac{D-1}{2}}\frac{a_k^2\mu_k^2}{r^2+a_k^2} \\
    =\sum_{k=1}^{\frac{D-1}{2}}\frac{r^2\mu_k^2}{r^2+a_k^2}.
\end{equation}
\subsection{D even, d odd case}
In the even $D$ case there is an additional coordinate $z=r\mu_0$ and a new variable 
 $\mu_0$, with the definition of $\mu_n$ such that $\mu_n^2= 1-\mu_0^2-\sum_{k=1}^{\frac{D-3}{2}}\mu_k^2$. All in all this only adds a row and a column easily included into the previous computation. The Jacobian matrix becomes
\begin{equation}
    {\cal J}_{ij}=\begin{pmatrix}
        \frac{\partial x_1}{\partial r} & \frac{\partial x_1}{\partial \mu_1} & \frac{\partial x_1}{\partial \phi_1} & 0 & 0 & ... & ...  & 0 & 0 \\
 \frac{\partial y_1}{\partial r} & \frac{\partial y_1}{\partial \mu_1} & \frac{\partial y_1}{\partial \phi_1} & 0 & 0 & ... & ...  & 0 & 0\\
 \frac{\partial x_2}{\partial r} & 0 & 0 & \frac{\partial x_2}{\partial \mu_2} & \frac{\partial x_2}{\partial \phi_2} & ... & ...  & 0 & 0\\
  \frac{\partial y_2}{\partial r} & 0 & 0 & \frac{\partial y_2}{\partial \mu_2} & \frac{\partial y_2}{\partial \phi_2} & ... & ...  & 0 & 0\\
  \vdots & \vdots & \vdots & \vdots & \vdots & \vdots & \vdots  & \vdots & \vdots\\
  \frac{\partial x_n}{\partial r} & \frac{\partial x_n}{\partial \mu_1} & 0 & \frac{\partial x_n}{\partial \mu_2} & 0 & ... & ...  & \frac{\partial x_n}{\partial \phi_n} & \frac{\partial x_n}{\partial \mu_0} \\
  \frac{\partial y_n}{\partial r} & \frac{\partial y_n}{\partial \mu_1} & 0 & \frac{\partial y_n}{\partial \mu_2} & 0 & ... & ...  & \frac{\partial y_n}{\partial \phi_n} & \frac{\partial y_n}{\partial \mu_0}\\
  \frac{\partial z}{\partial r} & 0 & 0 & 0 & 0 & 0 & 0  & 0 & \frac{\partial z}{\partial \mu_0}
    \end{pmatrix}\ .
\end{equation}
As in the odd $D$ case, we collect the first factor from all the rows. An extra factor of $\mu_{0}=\partial z/\partial r$  appears. We then expand with respect to the first column using Laplace method. For the terms of the expansion up to the second to the last it is enough to develop the last line before obtaining the same expansion as before. This only adds a factor
\begin{equation}
    \frac{\partial z}{\partial \mu_0}\Big(\frac{\partial z}{\partial r}\Big)^{-1}=\frac{r}{\mu_0}\ ,
\end{equation}
that, combined with the factor $\mu_0$ above, produces only an extra  $r$ factor that cancels a  ${1}/{r}$ factor. 
Moreover we have a term from the expansion of the determinant with respect to the last term in the first column, expanding the last column, in each of the two terms, we are left with a block-matrix determinant that yields
\begin{equation}
\begin{aligned}
    &\mu_0\Bigg(\prod_{l=1}^{\frac{D-2}{2}}\frac{\partial x_l }{\partial r }\frac{\partial y_l}{\partial r}\Bigg)\Bigg[\prod_{k=1}^{\frac{D-4}{2}}\Bigg(\frac{\partial x_k}{\partial \mu_k}\Big(\frac{\partial x_k}{\partial r}\Big)^{-1}\frac{\partial y_k}{\partial \phi_k}\Big(\frac{\partial y_k}{\partial r}\Big)^{-1}-\frac{\partial x_k}{\partial \phi_k}\Big(\frac{\partial x_k}{\partial r}\Big)^{-1}\frac{\partial y_k}{\partial \mu_k}\Big(\frac{\partial y_k}{\partial r}\Big)\Bigg)^{-1}\Bigg] \\
&\times\Bigg[\frac{\partial y_n}{\partial \mu_0}\Big(\frac{y_n}{r}\Big)^{-1}\frac{x_n}{\phi_n}\Big(\frac{x_n}{r}\Big)^{-1}-\frac{x_n}{\mu_0}\Big(\frac{\partial x_n}{\partial r}\Big)^{-1}\frac{\partial y_n}{\partial\phi_n}\Big(\frac{\partial y_n}{\partial r}\Big)^{-1}\Bigg].
\end{aligned}
\end{equation}
Including this term in the rest gives
\begin{equation}
    {\cal J}_{e}=(-1)^{D-2}\Bigg(\prod_{k\neq n}^{\frac{D-2}{2}}\mu_k\Bigg)\Pi \widetilde{F}+(-1)^{D-2}\mu_0^2\Bigg(\prod_{k=1}^{\frac{D-2}{2}}(r^2+a_k^2)\Bigg)\Bigg(\prod_{k\neq n}^{\frac{D-2}{2}}\mu_k\Bigg)\ ,
\end{equation}
or more the one for compactly
\begin{equation}
{\cal J}_{e}=\Bigg(\prod_{k\neq n}^{\frac{D-2}{2}}\mu_k\Bigg){F}\Pi\ ,
\end{equation}
where $F$ is defined as
\begin{equation}
    {F}=1-\sum_{k=1}^{\frac{D-2}{2}}\frac{a_k^2\mu_k^2}{r^2+a_k^2} 
    =\sum_{k=1}^{\frac{D-2}{2}}\frac{r^2\mu_k^2}{r^2+a_k^2}+\mu_0^2=\widetilde{F}+\mu_0^2\ .
\end{equation}
Notice that the Jacobian for $D$ even differs from the one for $D$ odd for the absence of the factor ${1}/{r}$.

\bibliographystyle{JHEP}
\bibliography{biblio}

\end{document}